\documentclass{aa}  

\usepackage{natbib}
\usepackage{graphicx}
\usepackage{txfonts}
\usepackage[colorlinks=true, allcolors=blue]{hyperref}
\begin{document} 

   \title{Prompt stellar and binary black hole mergers in tight triples}
   \subtitle{Insights from chemically homogeneous evolution}

   \author{A. Vigna-G\'omez
          \inst{1}
          \and
          E. Grishin\inst{2,3}
          \and J. Stegmann \inst{1}                    
          \and A. Olejak \inst{1}          
          \and S. A. Popa \inst{1}                    
          \and B. Liu\inst{4}
          \and A. S. Rajamuthukumar \inst{1}
          \and L. A. C. van Son\inst{5,6}
          \and A. Bobrick\inst{2,3}
          \and A. Dorozsmai\inst{7}
          }

   \institute{Max-Planck-Institut f\"ur Astrophysik, 
Karl-Schwarzschild-Str. 1, 85748 Garching, Germany\\ 
              \email{avigna@mpa-garching.mpg.de}
              \and
            School of Physics and Astronomy, Monash University, Clayton, VIC 3800, Australia
              \and
              OzGrav: Australian Research Council Centre of Excellence for Gravitational Wave Discovery, Clayton, VIC 3800, Australia
              \and
              Institute for Astronomy, School of Physics, Zhejiang University, 310058 Hangzhou, China
              \and
              Center for Computational Astrophysics, Flatiron Institute, New York, NY 10010, USA     
              \and
              Department of Astrophysical Sciences, Princeton University, 4 Ivy Lane, Princeton, NJ 08544, USA
              \and
              National Astronomical Observatory of Japan, National Institutes of Natural Sciences, 2-21-1 Osawa, Mitaka, Tokyo 181-8588, Japan
             }

   \date{Received ... ; accepted ...}
 
  \abstract 
    {
    Short-period massive binary stars are predicted to undergo a chemically homogeneous evolution (CHE), making them prime candidates for producing binary black holes (BBHs) that have the potential to merge within the age of the Universe. Most of these binaries have a tertiary companion and here we explore how a nearby third body could possibly influence this evolutionary channel. Our analysis combines analytic treatments of triple dynamics with insights from detailed stellar evolution models, focusing on the role of the von Zeipel-Lidov-Kozai mechanism, while also accounting for tidal and general relativistic apsidal precession. We examine the dynamics of triples at three critical evolutionary stages: the zero-age main sequence, shortly after the main sequence, and at the time of BBH formation. We find that, for triples with outer orbital periods less than 70 d(120 d), the inner binary can merge during(or after) the main sequence stage, leading to a hydrogen-rich (helium-rich) stellar merger. If a stellar merger is avoided, the inner binary could eventually form a BBH. In mildly hierarchical triples, with outer periods of around 100 d, the tertiary component can trigger a rapid merger of the BBH on timescales comparable to the outer orbital period. Stellar tides play a crucial role in determining the fate of the inner binary in such tight triple systems, as they can suppress the perturbative effects of the third star. When tidal forces damp the oscillations induced by the tertiary, the BBH merger may occur soon after stellar collapse. Notably, these outcomes are not restricted to CHE binaries but they can also be applied to any BBH formed from stars in tight orbits. Mergers in these systems are characterised by the proximity of a tertiary companion and the presence of recently ejected gas, making them promising candidates for electromagnetic counterparts and gravitational wave signals influenced by nearby tertiary objects.
   }

   \maketitle
%
%-------------------------------------------------------------------
\section{Introduction}
Stars that are several times more massive than the Sun typically form and evolve in multiple systems \citep[e.g.][]{2023ASPC..534..275O,2024NatAs...8..472L}, where they are orbited by one or more distant stellar companions.
In multiple-star systems, the evolution of stars can be influenced by the interplay of mass transfer episodes, explosions, and gravitational dynamics \citep[e.g.][]{2006epbm.book.....E,2023pbse.book.....T}.
Recent discoveries reveal that some systems initially classified as binaries are in fact part of a triple in which the most massive star is the outermost tertiary \citep[e.g.][]{2022MNRAS.511.4710E,2024ApJ...974...25K}.
In triples, the tertiary star can dynamically influence the evolution of the system.
In this paper, we examine the role of a tertiary star in `tight' high-mass stellar triples,  where the tertiary's outer orbital period is less than 1000 days. Our focus in this work is on binaries that experience chemically homogeneously evolution (CHE).

This type of evolution has been proposed in the context of single, rapidly rotating stars, where rotationally induced mixing leads to contraction, rather than the more typical expansion observed in non-rotating stars during the main sequence \citep[see][and reference therein]{1987A&A...178..159M}.
Overall, CHE is predicted to occur for (initially) massive stars ($\gtrsim 40\ M_{\odot}$) in short-orbital-period binaries ($\lesssim3\,\rm d$) at low metallicities ($Z \ll Z_{\odot}$), where tidal forces and minimal mass loss maintain the stars rotating rapidly \citep[e.g.][]{2009A&A...497..243D,2013A&A...556A.100S,2020A&A...641A..86H}.
In addition, CHE stars are the predicted progenitors of long gamma-ray bursts \citep{2006A&A...460..199Y,2018ApJ...858..115A}.
CHE binaries that remain massive and compact throughout their evolution are predicted to form binary black holes (BBHs) that can merge within the age of the Universe \citep{2016A&A...588A..50M,2016MNRAS.458.2634M,2016MNRAS.460.3545D,2020MNRAS.499.5941D,2021MNRAS.505..663R}, making them potentially detectable through gravitational waves.
CHE models predict that BBHs formed through this channel exhibit component masses  $\gtrsim 17\ M_{\odot}$, nearly equal-mass ratios, and moderate--to-high black-hole spins that are aligned with the orbital angular momentum \citep[e.g.][]{2020MNRAS.499.5941D,2024A&A...691A.339M}. 
Notably, the inferred properties of BBH merger GW190517 satisfy these criteria \citep{2022ApJ...941..179Q} and they are consistent with a potential CHE origin.
In addition, multiple stellar candidates have been proposed in the literature \citep[see][and references therein]{2024ApJ...966....9S}.
However, there is no confirmed detection of a CHE binary from observations of stellar binaries.

Given that $\sim 73\pm16\%$ of massive ($\gtrsim 16\ M_{\odot}$) main sequence stars are found in triples and/or quadruples \citep{2017ApJS..230...15M}, it is natural to consider CHE binary evolution with a tertiary companion. 
Tight hierarchical triples with CHE inner binaries were first studied semi-analytically in the context of gravitational-wave sources, where a coplanar triple may undergo two sequential BBH mergers that result in a peculiar, massive black-hole remnant \citep{2021ApJ...907L..19V}. 
The CHE scenario was subsequently integrated into rapid population synthesis \citep{2021MNRAS.505..663R}, enabling the exploration of its role within a stellar triple system \citep{2024MNRAS.527.9782D}.
The rapid population synthesis of triples with a CHE binary has shed light on two particular outcomes. 
First, mass transfer of the tertiary onto the inner binary is common, potentially affecting the evolution of the inner binary \citep{2024MNRAS.527.9782D,2025A&A...693A..84K}. 
Second, triple dynamics can dynamically assist a merger in the inner binary \citep{2024MNRAS.527.9782D}. 
A dynamically assisted merger can happen swiftly \citep[e.g.][]{2022MNRAS.515L..50V}, reducing the system from a triple to a binary \citep[e.g.][]{2024arXiv241214022P}.
This change impacts the number of tight inner massive binaries that would undergo CHE if they evolved in isolation.

In this paper, we present an analysis of the impact of a tertiary on dynamically assisted mergers in a CHE binary.
Specifically, we assess the role of the von Zeipel-Lidov-Kozai (ZLK) mechanism \citep{1910AN....183..345V, 1962P&SS....9..719L, 1962AJ.....67..591K} in tight triples. 
In the ZLK mechanism, coherent torques exchange the angular momentum between the inner and outer orbits of a hierarchical triple \citep{2016ARA&A..54..441N}.
The early studies of the ZLK mechanism were done in the test-particle limit, where one member of the inner binary is nearly massless, so that the outer angular momentum significantly exceeds the inner one.
In this limit, the mutual inclination required to induce a large inner eccentricity is close to 90 deg \citep[e.g.][and references therein]{2013MNRAS.431.2155N,2023MNRAS.522..937T}.
However, as the ratio of the inner to the outer orbital angular momentum increases, the critical inclination varies \citep{2017MNRAS.467.3066A, 2021MNRAS.500.3481H}. 
Systems with mild hierarchies, where the ratio of the inner and outer orbital period is not too small, can also achieve larger eccentricities \citep{2012ApJ...757...27A,2015MNRAS.452.3610A, 2016MNRAS.458.3060L, 2017MNRAS.466..276G, 2018MNRAS.481.4907G}.
Nonetheless, additional sources of apsidal precession, such as short range forces, may suppress any eccentricity excitation \citep{2011ApJ...741...82T,2015MNRAS.447..747L, 2017MNRAS.467.3066A, 2018MNRAS.474.3547G, 2020Natur.580..463G,2021MNRAS.502.2049L}. 
Recently, \cite{2022ApJ...934...44M} integrated these effects into a unified formula for the maximal eccentricity, applicable to any dynamically stable triple system with an equal-mass inner binary. 
In this study, we apply these results to identify orbital configurations that lead to either a stellar merger of the inner binary, or high eccentricities when the inner binary has reached a BBH configuration, resulting in a short coalescence time.

The structure of this paper is as follows.
First, we present a detailed stellar model of a CHE binary star system (Sect. \ref{sec:meth:binary_evolution}), which we integrate with our analytical method to assess the effect of ZLK oscillations (Sect. \ref{sec:meth:triple_dynamics}).
We proceed to analytically calculate the maximum eccentricity that a tertiary body can induce on the inner binary and evaluate how short-range forces might partially or completely suppress any increase in eccentricity (Sect. \ref{sec:res:maximum_eccentricity}).
We then determine which CHE binaries will merge as hydrogen-rich or helium-rich binaries (Sect. \ref{sec:res:fraction_of_mergers}) and discuss the outcome of BBH that survived stellar evolution (Sect. \ref{sec:res:effects_on_populations}).
We discuss our results and their implications (Sect. \ref{sec:discussion}). Finally, we present  a summary and our conclusions (Sect. \ref{sec:summary_and_conclusions}). 

% %-----------------------------------------------------------------

\section{Methods}\label{sec:meth}

\subsection{Binary evolution}\label{sec:meth:binary_evolution}

\subsubsection{Stellar models}\label{sec:meth:stellar_models}
We used the open-source one-dimensional stellar evolution code Modules for Experiments in Stellar Astrophysics (MESA; \citealt{2011ApJS..192....3P,2013ApJS..208....4P,2015ApJS..220...15P,2018ApJS..234...34P,2019ApJS..243...10P}), version 11701, to create a model of a CHE binary.
Our setup has identical physical assumptions to those reported in \cite{2020MNRAS.499.5941D}, which is closely aligned with the initial numerical implementation of CHE binaries presented in \cite{2016A&A...588A..50M}; however, the former study extends this process until core carbon depletion.
Our default CHE binary consists of equal-mass components on a circular orbit with inner orbital period $P_{\rm{in}}=1.1$ d, each with a zero-age main sequence (ZAMS) mass of $m_1=m_2=55\ M_{\odot}$.
The stellar models were constructed by placing two non-rotating stars in a close binary system. 
Once the system rapidly reached a synchronous state, we started simulating the system.
We chose an initial metallicity of $Z=0.00042$ (for reference, Solar metallicity is $Z_{\odot}=0.0142$, \citealt{2009ARA&A..47..481A}).
We  used this CHE model to evaluate the impact of a tertiary on the massive short-period inner binary.

\begin{figure}
    \centering
\includegraphics[width=0.5\textwidth]{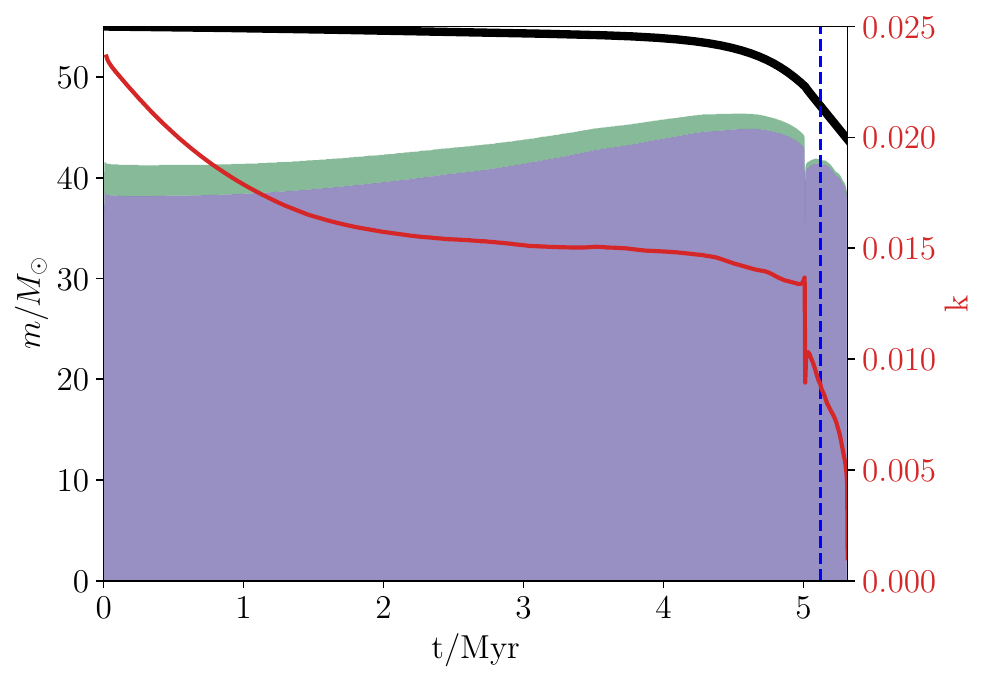}
\includegraphics[width=0.5\textwidth]{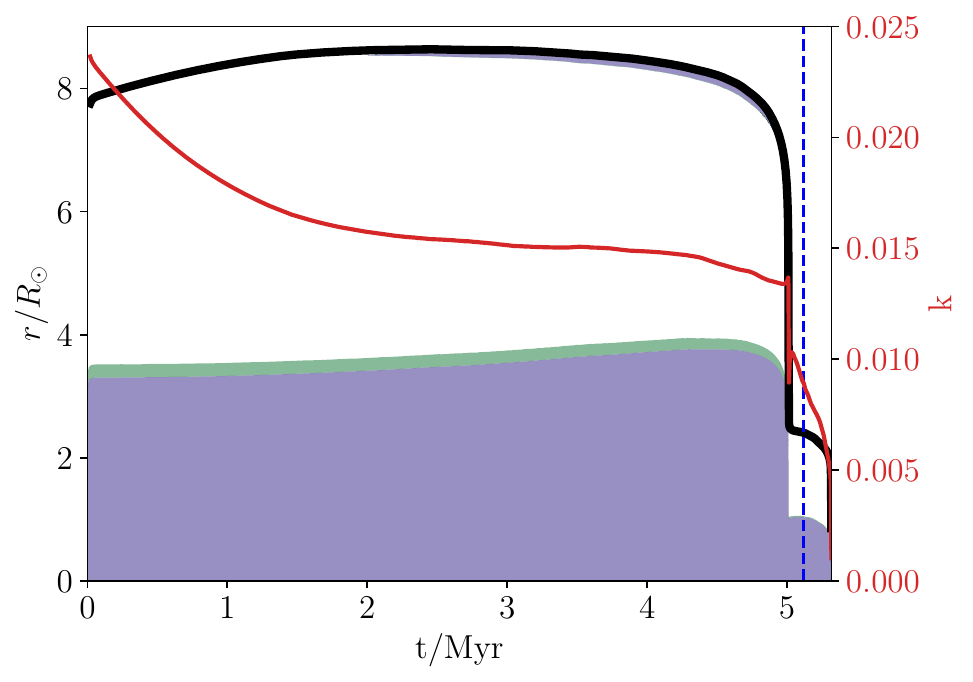}
    \caption{Kippenhahn diagrams showing mass ($m$) and radial ($r$) coordinates as a function of time ($t$) for a 55 $M_{\odot}$ star model with metallicity $Z=0.00042$, undergoing chemically homogenous evolution.
    The total mass or radius is shown with a thick solid black curve. 
    Convective regions are highlighted in purple, and regions with overshooting are marked in green.
    In addition, the apsidal motion constant ($k$) is shown as a solid red  line, with its scale in the right side of the ordinate axis.
    Finally, we show a vertical dashed blue line at 5.12 Myr; we later use this model to assess stellar mergers shortly after the end of the main sequence.
    }
    \label{fig:stellar_structure_low_Z}
\end{figure}

In Fig. \ref{fig:stellar_structure_low_Z}, we show the evolution of the internal structure of the stellar model, a graphic representation often referred to as a Kippenhahn diagram. 
In these Kippenhahn diagrams, we present the mass ($m$) and radial ($r$) coordinates as functions of time ($t$) for a component star in the CHE binary system, highlighting the evolution of the convective regions within these models.
In this CHE binary, both component stars evolve identically.
At the ZAMS, the $55 M_{\odot}$ star has a radius of $\approx7.6\ R_{\odot}$, and a convective core mass of $\approx37.8 M_{\odot}$ with a radius of $\approx3.5\ R_{\odot}$.
Throughout the main sequence, the star remains similarly compact, expanding up to $\approx 8.7\ R_{\odot}$.
By the end of the main sequence, the total mass has decreased to $\approx50 M_{\odot}$ while the convective core has grown in mass to $\approx46 M_{\odot}$.
After the main sequence, the star contracts to $\approx2.4\ R_{\odot}$, at which point the core temperature rises sufficiently to initiate the core-helium-burning phase.
Throughout this phase, and until the end of the evolution, the model slowly contracts while losing mass through stellar winds. 
In addition, this mass loss widens the orbit over time, leading to a final orbital period of $\approx$1.7 days by the end of the evolution.
Our last model has a final mass of $\approx43.8 M_{\odot}$ and a radius of $\approx0.9 R_{\odot}$.
For this orbital configuration, if we consider these systems as BBH progenitors, the time to coalescence through gravitational-wave emission would be $\approx450$ Myr \citep{1964PhRv..136.1224P}.

\subsubsection{Apsidal motion constant}
The apsidal motion constant is crucial for evaluating the impact of external precession on the inner orbit and its competition with the ZLK mechanism (Sect. \ref{sec:meth:analytics}).
Figure \ref{fig:stellar_structure_low_Z} illustrates the apsidal motion constant ($k$) as calculated in MESA. 
This dimensionless number is used to parametrise the internal structure of a star and is related to the Love number ($k_{\rm{L}}$) through the relationship $2k = k_{\rm{L}}$.
A lower $k$ value indicates a more centrally concentrated mass distribution of the star, which results in a slower rate of apsidal motion.
For our stellar model, the value of the apsidal motion constant at the ZAMS is $k\approx 0.024$; by the end of the main sequence, this value drops to $k\approx 0.015$.
After the main sequence, this value keeps decreasing; for the last saved model in our simulation $k\approx 0.0007$.

\subsection{Triple dynamics}\label{sec:meth:triple_dynamics}

\begin{figure}
    \centering
    % [trim={left bottom right top},clip]
    \includegraphics[trim={0.7cm 8.5cm 0.7cm 9.5cm},clip,width=\columnwidth]{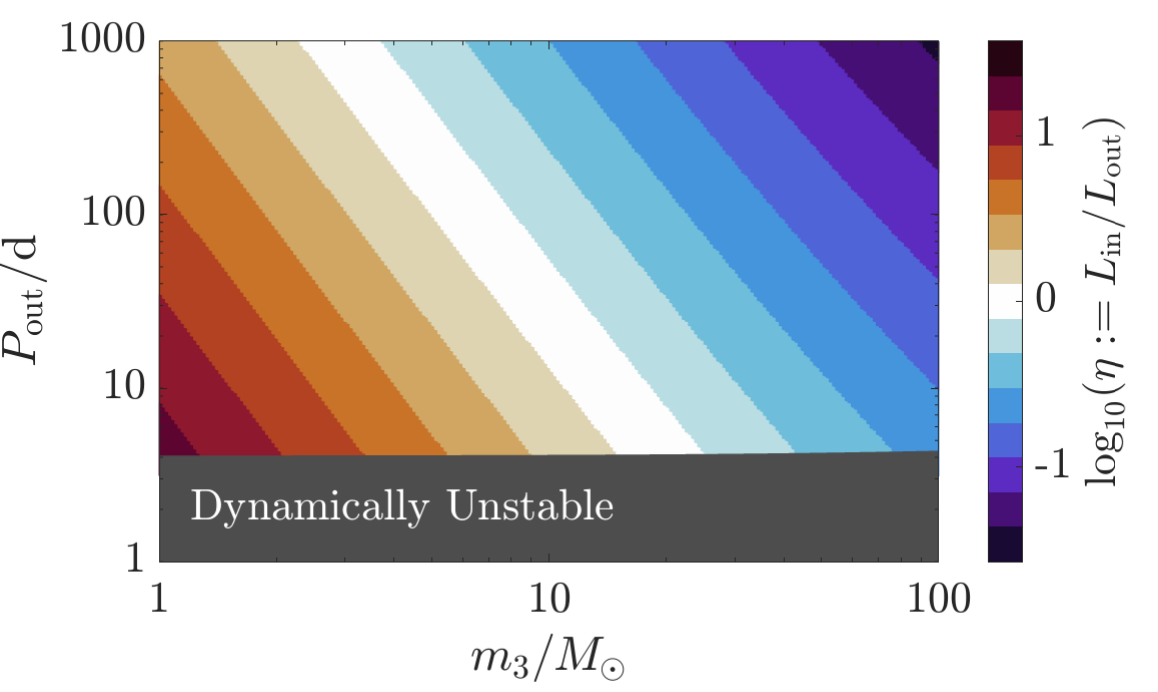}
    \includegraphics[trim={0.5cm 8.5cm 0.9cm 9.5cm},clip,width=\columnwidth]{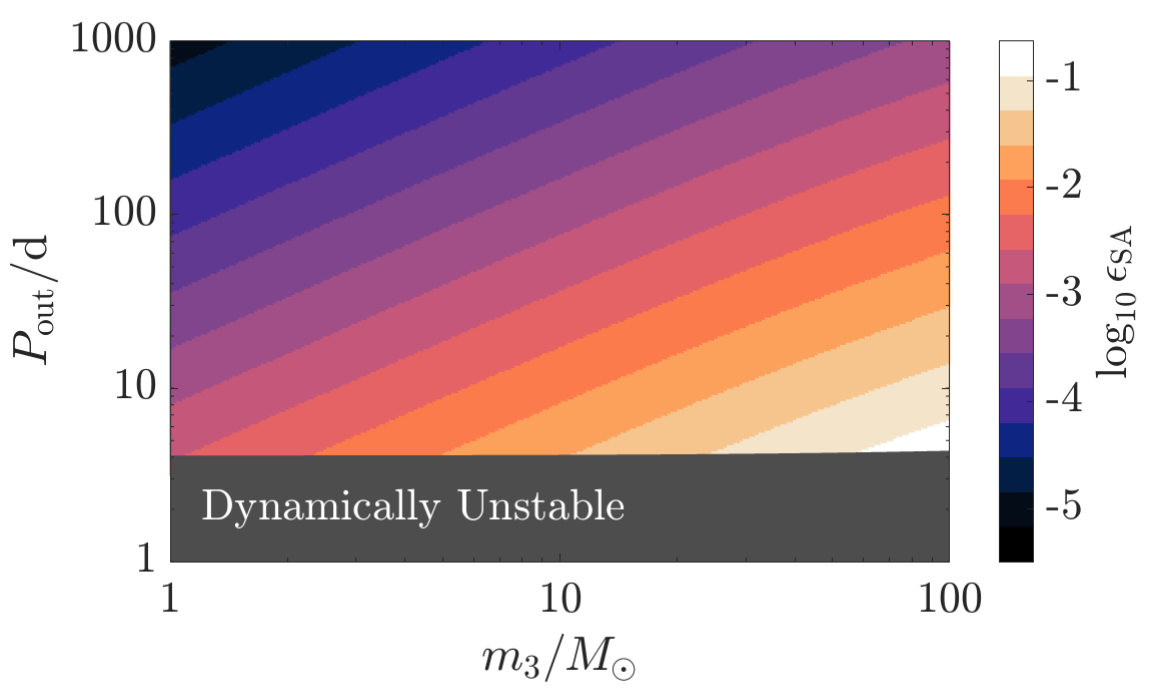}
    \caption{Dimensionless quantities for a triple system with an arbitrary tertiary mass ($m_3$) and outer orbital period ($P_{\rm out}$).
    The system includes an inner circular binary with masses $m_1=m_2=55\ M_{\odot}$ and an orbital period of 1.1 d, at a metallicity $Z=0.00042$, evaluated at the ZAMS.
    The dark grey region indicates dynamically unstable triples.
    We display the inner to outer orbital angular momentum ratio ($\eta := L_{\rm{in}}/L_{\rm{out}}$, top panel) and the single-averaging parameter ($\epsilon_{\rm SA}$, bottom panel).
    }
    \label{fig:adim_quantities}
\end{figure}

\begin{figure}
    \centering
    % [trim={left bottom right top},clip]
    \includegraphics[trim={0.7cm 8.5cm 0.7cm 9.5cm},clip,width=\columnwidth]{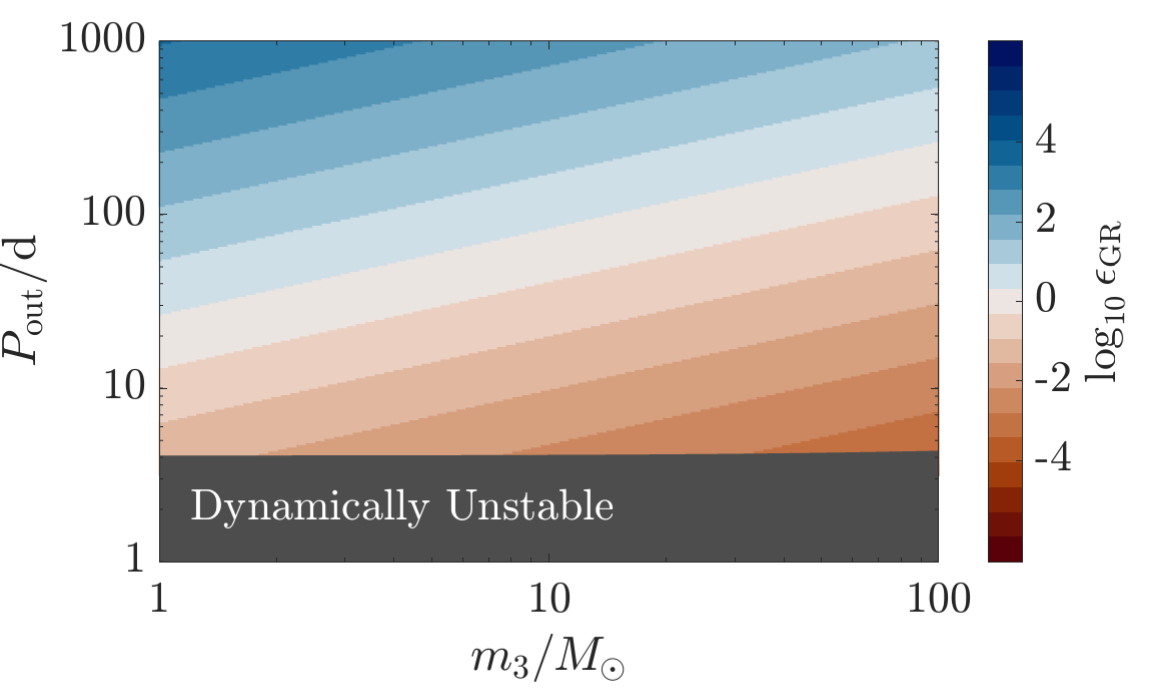}
    \includegraphics[trim={0.5cm 8.5cm 0.9cm 9.5cm},clip,width=\columnwidth]{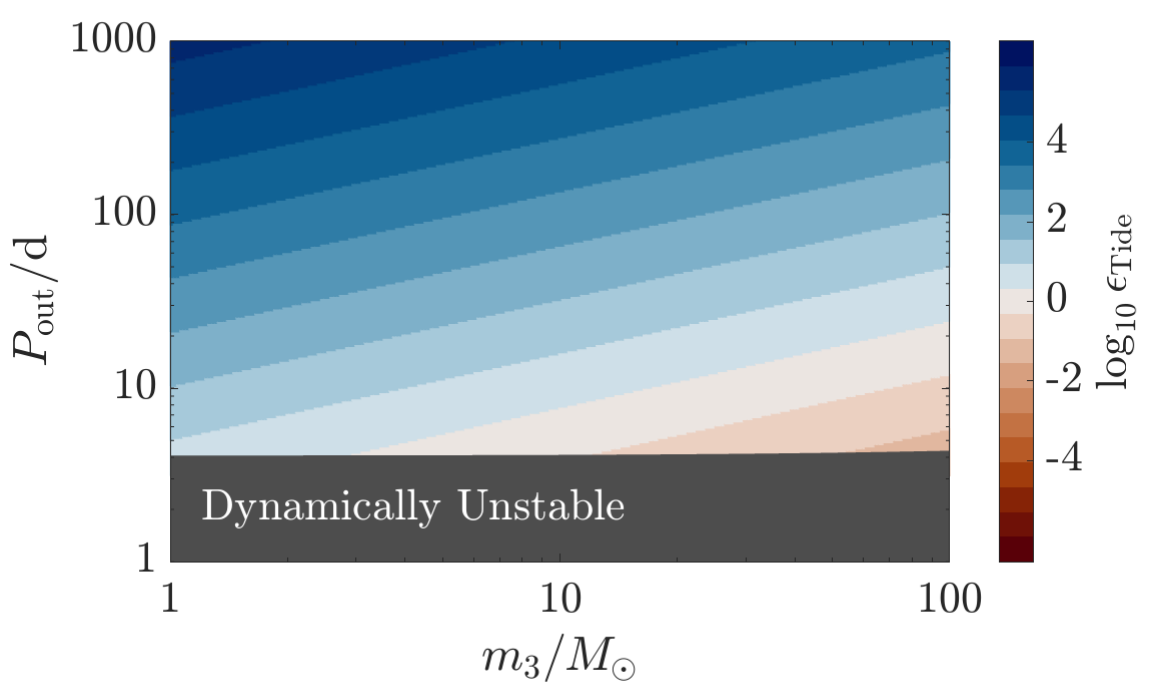}
    \caption{Dimensionless parameters quantifying the role of short-range forces for a triple system with an arbitrary tertiary mass ($m_3$) and outer orbital period ($P_{\rm out}$). 
    The system includes an inner circular binary with masses $m_1=m_2=55\ M_{\odot}$ and an orbital period of 1.1 d, at a metallicity of $Z=0.00042$, evaluated at the ZAMS.
    We use different colours to highlight various regions of interest.
    Dark grey indicates dynamically unstable triples.
    We display the relative strength of the apsidal advance due to GR ($\epsilon_{\rm GR}$, top panel) and the relative strength of the apsidal advance due to tides ($\epsilon_{\rm Tide}$, bottom panel).
    For a direct comparison, we select an identical scale for the colourbars. 
    }
    \label{fig:short_range_forces}
\end{figure}

\subsubsection{Assumptions}\label{sec:meth:assumptions}

To evaluate the impact of a tertiary companion on the inner binary, we used analytical formulae to determine dynamical stability and assess the role of ZLK oscillations.

The dynamical instability threshold is determined from \cite{2022MNRAS.516.4146V}. 
For circular inner and outer binaries, with eccentricities $e_{\rm in}=0$ and $e_{\rm out}=0$, it can be simplified to 
\begin{equation}
    a_{\rm out} = 2.4 (1+q_{\mathrm{out}})^{2/5} \left( \frac{\cos{i_0} - 1}{8} + 1 \right) a_{\rm in},
\label{eq:stability}
\end{equation}
where the subscripts `in' and `out' refer to the inner and outer orbits, $a_{\rm{in}}$ and $a_{\rm{out}}$ are the semi-major axes, $q_{\rm out} = (m_1+m_2)/m_3 \in [10^{-2}, 10^2]$ is the mass ratio, with $m_1$ and $m_2$ being the component masses of the inner binary and $m_3$ being the mass of the tertiary companion, and $i_0$ the initial mutual inclination.
In order to determine dynamical stability, we specifically considered the case where $i_0=0\ \rm{deg}$, as this condition results in the widest stable configurations.
Our stability criteria is closely aligned  with other works in the literature \citep{2001MNRAS.321..398M,2022PASA...39...62T}.
For our default CHE binary, we find that triples with $P_{\rm out}\lesssim 4$ d are dynamically unstable.
A dynamically unstable triple system will undergo a chaotic evolution, potentially involving component exchanges, mergers, or ejections \citep[e.g.][]{1994MNRAS.267..161K,1999ApJ...511..324I,2012ApJ...760...99P,2022A&A...661A..61T}. 
Predicting the outcome of this highly non-linear behaviour is beyond the scope of our analysis.

We assessed the role of ZLK oscillations following the analytic formulae from \cite{2022ApJ...934...44M}.
That study provides solutions that were tested against direct $N$-body integrations.
In particular, we sought to determine, for the given triple configurations,  the maximum eccentricity that the inner binary ($e_{\rm{max}}$) can attain given an initial mutual inclination. 
When using this formalism, we assume the following:
\begin{itemize}
    \item The inner binary is initially circular ($e_{\rm{in}}=0$) and has a mass ratio ($q_{\rm{in}}=m_2/m_1$) close to unity, which is the case for our CHE binary.
    \item The outer eccentricity is always circular ($e_{\rm{out}}=0$). In principle, our analysis can be extended for $e_{\rm{out}}\ne 0$ if the mass ratio is unity due to a gauge-freedom of the Hamiltonian forms, which keeps it axisymmetric \citep{2023MNRAS.522..937T, 2024MNRAS.533..486G}. Non-equal mass binaries with $e_{\rm{out}} \ne 0$ are susceptible to chaotic evolution and the analytical prescriptions below are not valid \citep[e.g.][]{2013MNRAS.431.2155N,2025arXiv250105506S}.
    \item We ignored the oblateness due to stellar rotation. 
    The additional force rotation induces on the inner binary is subdominant to the effect of the tidal bulge in synchronous systems. 
    Specifically, the extra precession induced from tidal bulges is $\sim30$ times larger than from rotation-induced oblateness, even for pseudo-synchronous binaries, where there is no net angular momentum flow \citep{2015MNRAS.447..747L}.
\end{itemize}

\subsubsection{Analytic formulae}\label{sec:meth:analytics}

To determine the maximal eccentricity ($e_{\rm max}$) of the inner binary in a triple system, we used the (implicit) analytic formulae in Eq. (17) and (24) from \cite{2022ApJ...934...44M}, which are briefly revised here. First, we define the inner ($L_{\rm in}$) to outer ($L_{\rm out}$) angular momentum ratio,
\begin{align}
\eta :=  \frac{L_{\rm in}}{L_{\rm out}} = \frac{\mu_{\rm in}}{\mu_{\rm out}}\Bigg[ \frac{(m_1+m_2)a_{\rm in}}{(m_1+m_2+m_3)a_{\rm out}(1 - e_{\rm out}^2)}\Bigg]^{1/2},
\label{eq:eta}
\end{align}
where $\mu_{\rm in} = m_1 m_2/(m_1+m_2)$  and $\mu_{\rm out} = (m_1+m_2) m_3/(m_1+m_2+m_3)$ are the reduced masses of the inner and outer binaries, respectively.
The value of $\eta$ is useful to quantify the hierarchy of the triple star system.
In Fig. \ref{fig:adim_quantities}, we show $\eta$ while varying the tertiary mass between $1\leq m_3/M_{\odot} < 100$ and the outer orbital period between $1 < P_{\rm out}/\rm{d} < 1000$, which defines our parameter-space region of interest.
As expected, higher-mass tertiaries tend to be in the test particle limit ($\eta \ll 1)$. 
In contrast, low tertiary masses $m_3 \lesssim 10\ M_{\odot}$ can result in the orbital angular momentum of the inner binary dominating over that of the outer orbit ($\eta > 1$).

The analysis of triples is often simplified using the double-averaged (DA) approximation. 
In this approach, the gravitational effects on both the inner and outer orbits are averaged over one orbital period. 
Alternatively, the single-averaged (SA) approximation averages only the inner orbit over time. 
We define the dimensionless SA parameter as
\begin{equation}
    \epsilon_{\rm SA} := \left[\frac{a_{\rm{in}}}{a_{\rm{out}}(1-e_{\rm{out}}^2)}\right]^{3/2} \left[ \frac{m_3^2}{(m_1+m_2)(m_1+m_2+m_3)} \right ]^{1/2}  \label{eps_sa}
,\end{equation}
which allows us to correct the double averaged approximation \citep{2016MNRAS.458.3060L}.
In Fig. \ref{fig:adim_quantities} we also show $\epsilon_{\rm SA}$ in our parameter-space region of interest. 
For hierarchical triples, $\epsilon_{\rm SA}\lesssim 0.25$, and the DA approximation is well justified for $\epsilon_{\rm SA}\ll 1$.

We proceeded to assess the relative importance of short-range forces, which can also be encapsulated with dimensionless parameters. 
We followed the definitions of \cite{2015MNRAS.447..747L} and write the relative strength of the apsidal precession due to general relativity (GR) and tides as
\begin{equation} 
\epsilon_{\textrm{GR}} := \frac{3(m_1+m_2)}{m_3}\Bigg(\frac{a_{\rm{out}}}{a_{\rm{in}}}\Bigg)^3\frac{r_g}{a_{\rm{in}}}
\end{equation}
and
\begin{equation}\label{eq:epsilon_tide}
\quad \epsilon_{\textrm{Tide}} := 2\times \frac{15m_1(m_1+m_2)(2k) a_{\rm{out}}^3 R_c^5}{m_2m_3 a_{\rm{in}}^8},
\end{equation}
respectively, where $r_g=G(m_1+m_2)/c^2$ is the combined Schwarzschild radius and $R_c$ is the convective radius of the star, which is generally smaller than the stellar surface ($R$). 
These parameters, $\epsilon_{\rm GR}$ and $\epsilon_{\rm Tide}$, quantify the relative importance of the short-range potential terms with respect to the quadruple potential leading to the ZLK effect \citep{2015MNRAS.447..747L}.
For small $\epsilon_{\rm GR} \ll 1$ and $\epsilon_{\rm Tide} \ll 1$ the orbital evolution is approximated by Newtonian point masses and the eccentricity determined exclusively by the initial mutual inclination.
For large values $\epsilon_{\rm GR} \gg 1$ and $\epsilon_{\rm Tide} \gg 1$, additional parameters become relevant and the eccentricity evolution may be suppressed. Consequently, the binary system may evolve as if isolated, unaffected by the tertiary component.

In Fig. \ref{fig:short_range_forces} we show these dimensionless parameters for our default CHE binary (Sect. \ref{sec:meth}) at the ZAMS.
In the top and bottom panels we display $\epsilon_{\rm GR}$ and $\epsilon_{\rm Tide}$, respectively.
The dynamic range of these two parameters is broad, spanning several orders of magnitude across our triple orbital parameter space of interest.
We find that, among short-range forces, tidal effects dominate the evolution at the ZAMS. 

The values of $\epsilon_{\rm SA}$, $\epsilon_{\rm GR}$, and $\epsilon_{\rm Tide}$ are directly proportional to the potential energy each assumption or short-range force contributes to the complete Hamiltonian. 
They also enable us to derive an analytical solution for the maximum eccentricity that an inner binary can achieve through the ZLK effect.
With the inclusion of the tidal term,  Eq. (17) of \cite{2022ApJ...934...44M} is modified to
\begin{align}
    0=&-5g_{\rm max}^2 + j_{\rm min}^2\Bigg(3 - \eta\cos i_0 -\frac{\eta^2e_{\rm max}^2}{4} \Bigg) \nonumber \\
    &+ \frac{3\epsilon_{\rm SA}}{8}\Bigg[ -\frac{j_{\rm min}^2\cos^3i_0}{e_{\rm max}^2} - \frac{\eta j_{\rm min}^2}{2} -9j_{\rm min}^2 g_{\rm max} \nonumber \\ 
    &+g_{\rm max}^3\Bigg(\frac{1}{e_{\rm max}^2} - 16\Bigg)  \Bigg] +  \frac{8\epsilon_{\rm GR}j_{\rm min}( j_{\rm min} - 1)}{3e_{\rm max}^2} \nonumber \\ 
    &+ \frac{8\epsilon_{\rm Tide}}{45} \left[1 - \frac{1 + 3e_{\rm max}^2 + (3/8)e_{\rm max}^4}{(1-e_{\rm max}^2)^{9/2}}\right],\label{eq:emax}
\end{align}
where $g_{\rm max}=\cos i_0 + \eta e_{\rm max}^2/2$, and $j_{\rm min}=\sqrt{1-e_{\rm max}^2}$. Eq.~\eqref{eq:emax} can be solved for $e_{\rm max}$ in terms of $\epsilon_{\rm SA}, \eta, i_0$, $\epsilon_{\rm GR}$, and $\epsilon_{\rm{tide}}$, which are determined from initial conditions.

In addition, there is a fluctuating term given by Eq. (24) in \citealp{2022ApJ...934...44M} as
\begin{equation}
    \delta{e} = \mathcal{C}\epsilon_{\rm SA} e_{\rm max}\left( j_{\rm min} - \frac{e_{\rm max}^2 }{2} \mathcal{C}\epsilon_{\rm SA} \right),
\end{equation}
where $\mathcal{C}$ is a constant of order unity defined in Eq. (23) in \cite{2022ApJ...934...44M}. The total eccentricity is then given by

\begin{equation}
    e=e_{\rm max} + \delta e.
\end{equation} 

We note that \cite{2022ApJ...934...44M} considers highly eccentric regimes.
In our analysis, we  ultimately found that the latter term for $\delta e$ is not significant during the stellar lifetime.
However, it may become relevant after the formation of a BBH, once tidal effects are no longer present.

\subsubsection{Merger criteria}\label{sec:meth:merger_criteria}

The ZLK mechanism results in a periodic exchange between the mutual inclination and the eccentricity of the inner binary.
However, due to conservation of angular momentum, the semi-major axis of the inner binary remains constant.
While the semi-major axis remains constant, changes in eccentricity affect the pericentre distance $r_{\rm p}=a(1-e)$.

For a stellar binary, the pericentre is crucial in order to determine whether or not a binary remains detached or if at least one components fills their Roche lobe.
For Roche-lobe overflow to begin, the radius of the star must be larger than the Roche radius ($R_{\rm RL}$).
Some binary systems, sometimes referred to as contact binaries, can have both stars filling their Roche lobes while remaining in a dynamically stable configuration \citep{2015ApJ...812..102A}. 
CHE binaries are predicted to exist in such contact configurations \citep{2016A&A...588A..50M}.

We follow the criteria from \cite{2016A&A...588A..50M} in order to determine whether or not a contact binary will lead to a merger.
In this criteria, if both stars overflow their outer Lagrangian point ($L_2$), the mass lost through this point will carry high specific angular momentum, leading to a stellar merger.
The criterion $R>L_2$ can be rewritten in terms of the Roche radius using $L_2 \approx 1.32 R_{\rm RL}$.
We calculate the Roche radius at pericentre as $R_{\rm RL}=r_{\rm p}f(q)$, where $f(q)$ is a function that is dependent on the mass ratio.
We utilised the fitting formula from \cite{1983ApJ...268..368E}, based on a circular orbit, to determine $f(q)$.
For a mass ratio of unity, as in our default CHE binary, the criterion is approximately simplified to
\begin{equation}
    R>0.5r_{\rm p},
\end{equation} 
and a merger occurs if the stars make contact at pericentre.

\begin{figure}
    \centering
    % [trim={left bottom right top},clip]
    \includegraphics[trim={0.7cm 8.5cm 0.7cm 9.5cm},clip,width=\columnwidth]{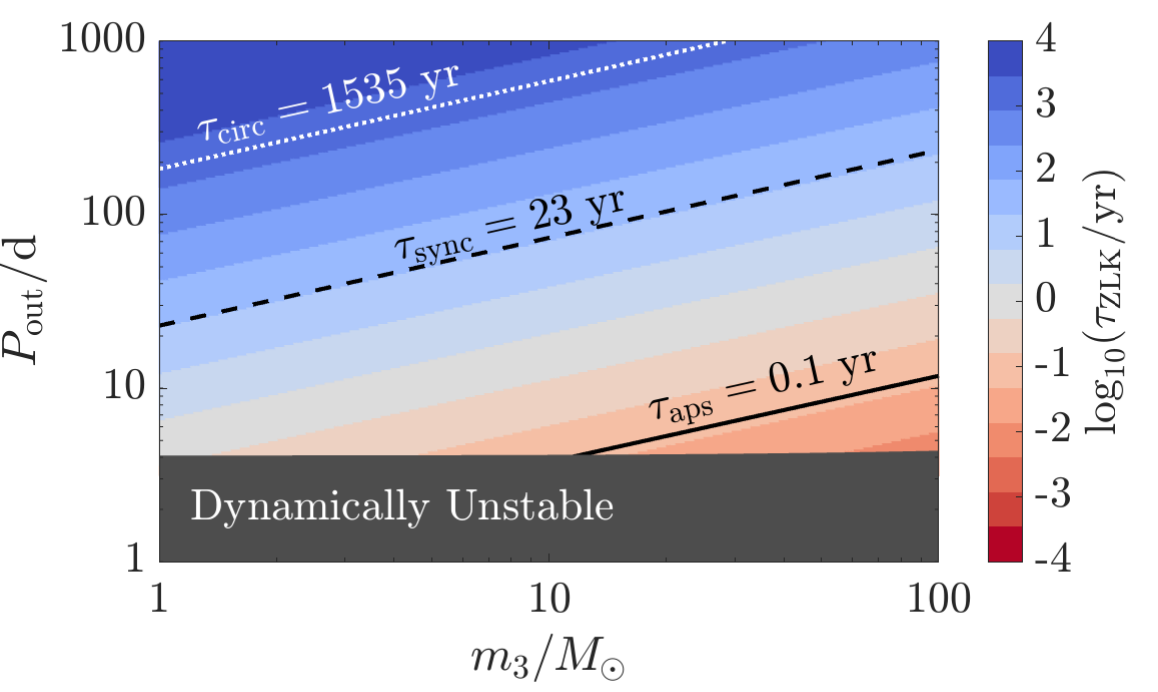}
    \caption{Characteristic ZLK timescale ($\tau_{\rm ZLK}$) for a triple system with an arbitrary tertiary mass ($m_3$) and outer orbital period ($P_{\rm out}$). 
    The system includes an inner circular binary with masses $m_1=m_2=55\ M_{\odot}$ and an orbital period of 1.1 d, at a metallicity of $Z=0.00042$, evaluated at the ZAMS. 
    The lines indicate different relevant timescales for the inner binary.
    The dotted white, dashed black, and solid black lines indicate the circularisation ($\tau_{\rm circ}$), synchronisation ($\tau_{\rm sync}$), and apsidal tidal ($\tau_{\rm aps}$) timescales, respectively.
    Dark grey indicates dynamically unstable triples.
    }
    \label{fig:timescales}
\end{figure}

\subsection{Timescales}\label{sec:meth:tides}

\subsubsection{ZLK timescale}
To estimate the timescale over which the ZLK mechanism can influence the eccentricity of the inner binary, we follow \cite{2015MNRAS.447..747L} and define the characteristic ZLK timescale\footnote{Equation \eqref{eq:zlk_timescale} is an order-of-magnitude estimate of the secular timescale, which is used to measure the different strength of the terms in the total Hamiltonian \citep{2015MNRAS.447..747L}. The more accurate estimate requires integrating the action $G_{\rm in} \propto \sqrt{1-e_{\rm in}^2}$ over a closed orbit in phase space \citep{2015MNRAS.452.3610A,2024MNRAS.533..486G}, which differs by an order unity from Eq. \eqref{eq:zlk_timescale}.} as
\begin{equation}\label{eq:zlk_timescale}
    \tau_{\rm ZLK} = \frac{(m_1+m_2)^{1/2}a_{\rm out}^{3}}{G^{1/2}m_3 a_{\rm in}^{3/2}}.
\end{equation}
In Fig, \ref{fig:timescales} we show $\tau_{\rm ZLK}$ in our parameter-space region of interest.
The largest values for the ZLK timescale are $\tau_{\rm ZLK}\sim 10^4$ yr, which is significantly shorter than the lifetime of the component stars of the binary ($\approx 5$ Myr; see Fig. \ref{fig:stellar_structure_low_Z}).
We proceed to estimate the relevant tidal timescales, which are solely determined by the structure of the inner binary, in order to compare them with the ZLK timescale, which depends on the triple orbit.

\subsubsection{Apsidal tidal timescale}
In a short-period binary system where a star is tidally deformed, a non-dissipative tidal bulge may form, resulting in a permanent quadrupole moment that influences the overall potential of the binary system.
This static tide induces additional precession \citep{2015MNRAS.447..747L}.
We follow \citealt{2015MNRAS.447..747L} and define the apsidal tidal timescale as
\begin{equation}
    \tau_{\rm aps} := \frac{\tau_{\rm ZLK}}{\epsilon_{\rm Tide}}.
\end{equation}

For our default CHE binary at the ZAMS (Sect. \ref{sec:meth:stellar_models}), this results in an apsidal tidal timescale of $\tau_{\rm{aps}} \approx 0.1\ \rm yr$.
By definition, the region where $\tau_{\rm ZLK} \approx \tau_{\rm aps}$ corresponds to $\epsilon_{\rm Tide} \approx 1$, which delimits the region where the static tide becomes strong enough to suppress the ZLK mechanism  ($\epsilon_{\rm Tide} \gtrsim 1$, Fig. \ref{fig:short_range_forces}).

\subsubsection{Tidal dissipation timescale}
In a stellar binary system, tidal dissipation eventually aligns both stellar spins with the orbit, synchronises the components with the orbital motion, and makes the orbit circular \citep[e.g.][and references therein]{2008EAS....29...67Z}. 
The efficiency of tidal dissipation, however, is highly dependent on the specific dissipation mechanism \citep{1977A&A....57..383Z} and is also very sensitive to the assumptions made about the stellar structure \citep[e.g.][]{2013A&A...550A.100S,2018A&A...616A..28Q}. 
In binary star systems with stars that have radiative envelopes, such as those in our CHE binary, the primary dissipation mechanism is through the dynamical tide \citep{1975A&A....41..329Z,1977A&A....57..383Z}. 

We follow \cite{2002MNRAS.329..897H} and define the circularisation and synchronisation timescales via the dynamical tide for a near-circular binary as 
\begin{equation}\label{eq:dynamical_tide_circ}
    \frac{1}{\tau_{\rm{circ}}} = \frac{21}{2}\Bigg( \frac{GM}{R^3} \Bigg)^{1/2} q_{\rm in}(1+q_{\rm in})^{11/6} E_2 \Bigg( \frac{R}{a_{\rm in}} \Bigg)^{21/2}
\end{equation}
and 
\begin{equation}\label{eq:dynamical_tide_sync}
    \frac{1}{\tau_{\rm{sync}}} = 5 \times2^{5/3}  \Bigg( \frac{GM}{R^3} \Bigg)^{1/2} \frac{1}{\beta^2} q_{\rm in}^2(1+q_{\rm in})^{5/6} E_2 \Bigg( \frac{R}{a_{\rm in}} \Bigg)^{17/2},
\end{equation}
where $M$ and $R$ are the mass and radius of the star, respectively, $E_2$ is the second-order tidal coefficient, used in the form
\begin{equation}
    E_2 = 1.592\times 10^{-9} (M/M_{\odot})^{2.84},
\end{equation}
and $\beta$ is the radius of gyration, which near the ZAMS we fit to
\begin{equation}
    \beta \approx 0.7k^{0.224},
\end{equation}
following the models from \cite{1989A&AS...81...37C}. For our default CHE binary at the ZAMS (Sect. \ref{sec:meth:stellar_models}), this results in a circularisation timescale of $\tau_{\rm{circ}} \approx 1500\ \rm yr$ and a synchronisation timescale of $\tau_{\rm{sync}} \approx 23\ \rm yr$.

\subsubsection{Comparison}\label{sec:meth:tides:comp}
In Fig. \ref{fig:timescales} we can see how the ZLK timescale compares to the apsidal tidal timescale (solid black line) and the tidal dissipation timescales specifically, the circularisation (dotted white line) and synchronisation (dashed black line) timescales.
We differentiate between four main regimes:
\begin{itemize}
    \item When $\tau_{\rm circ} \lesssim \tau_{\rm ZLK}$, tidal dissipation effectively quenches ZLK oscillations.
    \item When $\tau_{\rm sync} \lesssim \tau_{\rm ZLK} \lesssim \tau_{\rm circ}$, tidal dissipation can partially suppress ZLK oscillations.
    \item When $\tau_{\rm aps} \lesssim \tau_{\rm ZLK} \lesssim \tau_{\rm sync}$, tidal dissipation is negligible, but the static tide from tidal bulges partially suppresses the ZLK mechanism.
    \item When $\tau_{\rm ZLK} < \tau_{\rm aps}$, the ZLK mechanism is highly efficient.
\end{itemize}
We  explicitly demonstrate the impact of short-range forces and the ZLK mechanism on the orbital configuration of the inner binary in the next section.

\section{Results}
\label{sec:results}

\begin{figure*}
    \centering
    % [trim={left bottom right top},clip]
    \includegraphics[trim={0.7cm 8.5cm 0.7cm 9.5cm},clip,width=\columnwidth]{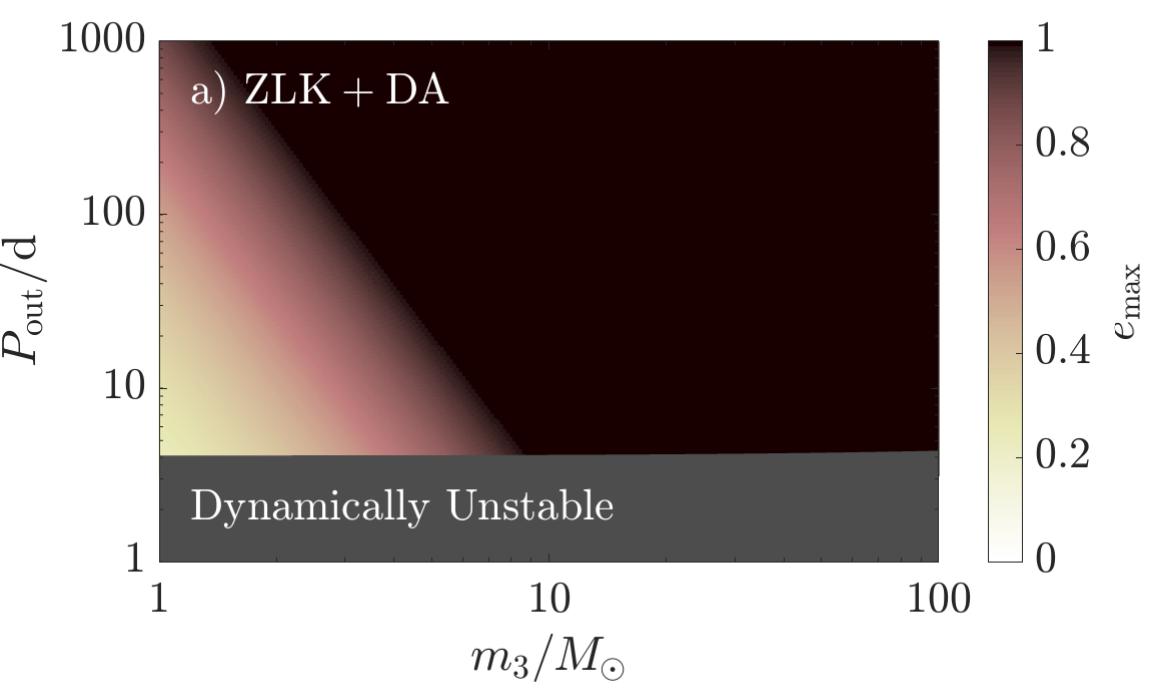}
    \includegraphics[trim={0.5cm 8.5cm 0.9cm 9.5cm},clip,width=\columnwidth]{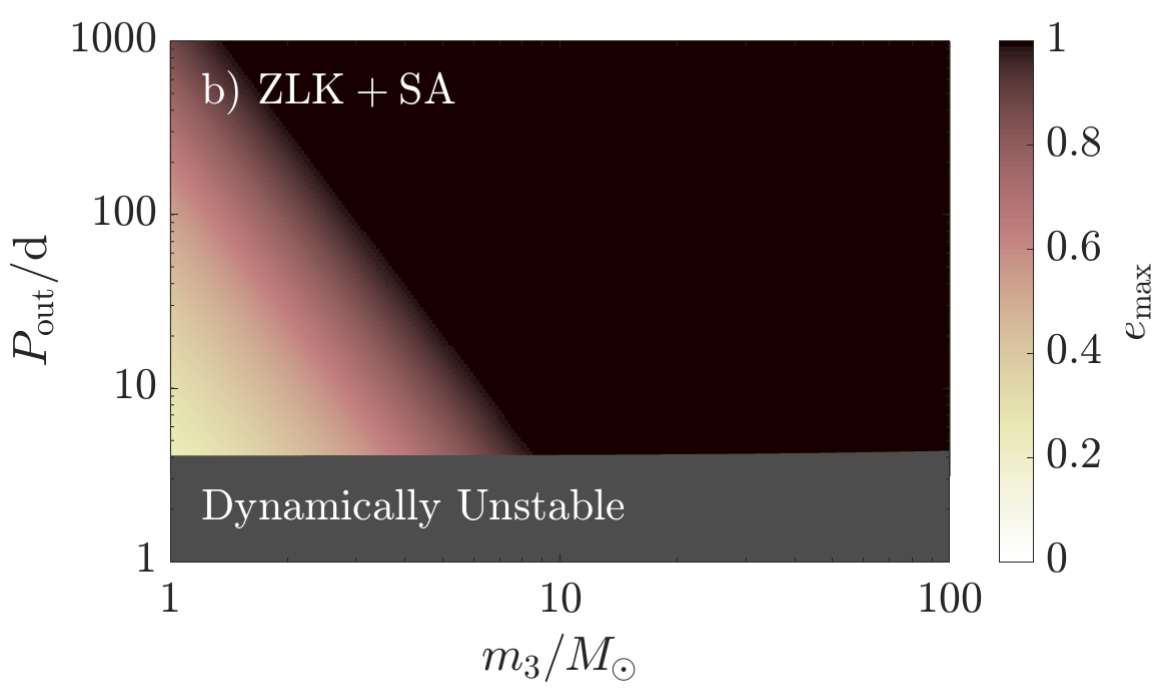}
    \includegraphics[trim={0.7cm 8.5cm 0.7cm 9.5cm},clip,width=\columnwidth]{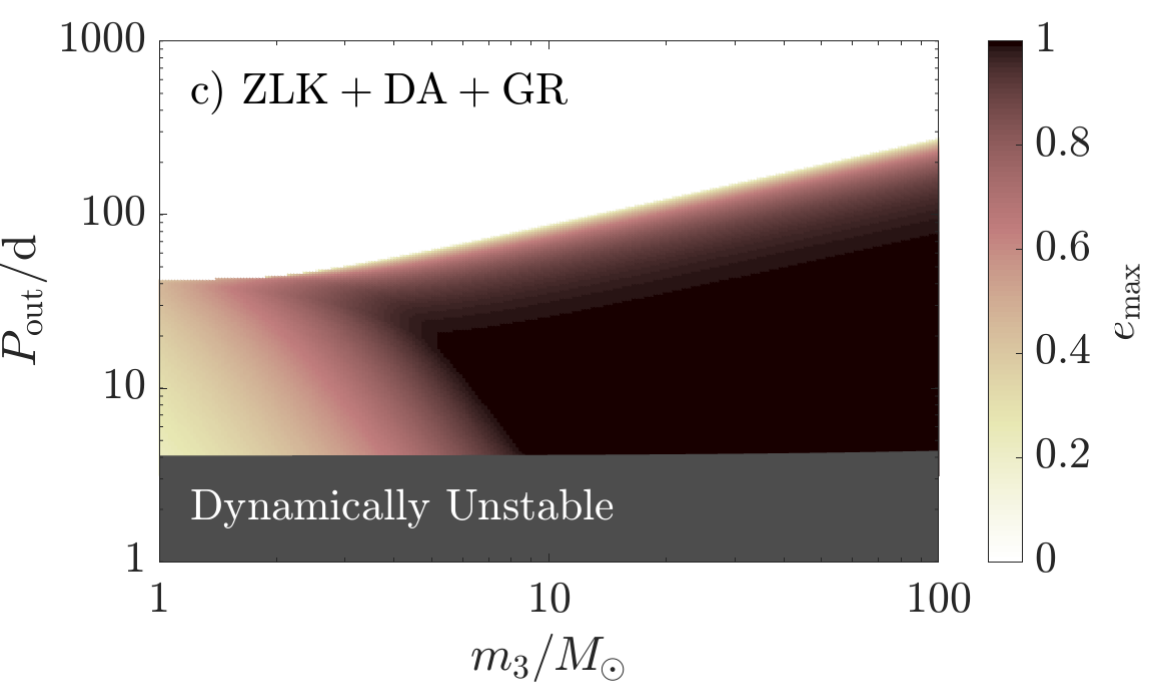}
    \includegraphics[trim={0.7cm 8.5cm 0.7cm 9.5cm},clip,width=\columnwidth]{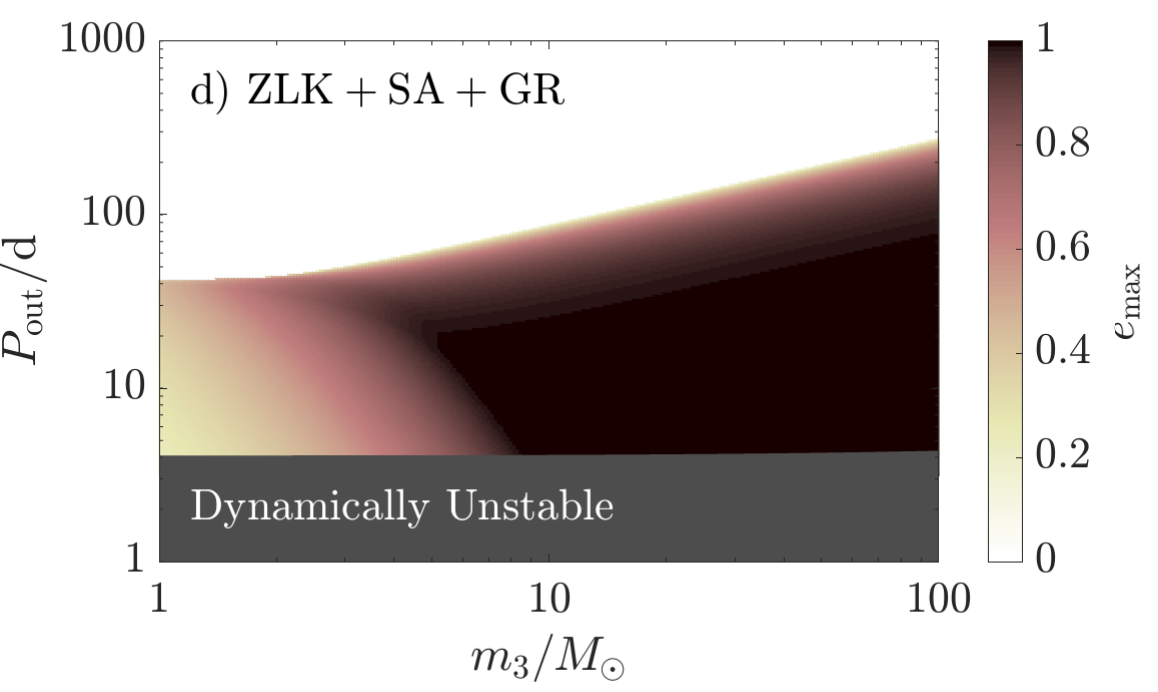}
    \includegraphics[trim={0.5cm 8.5cm 0.9cm 9.5cm},clip,width=\columnwidth]{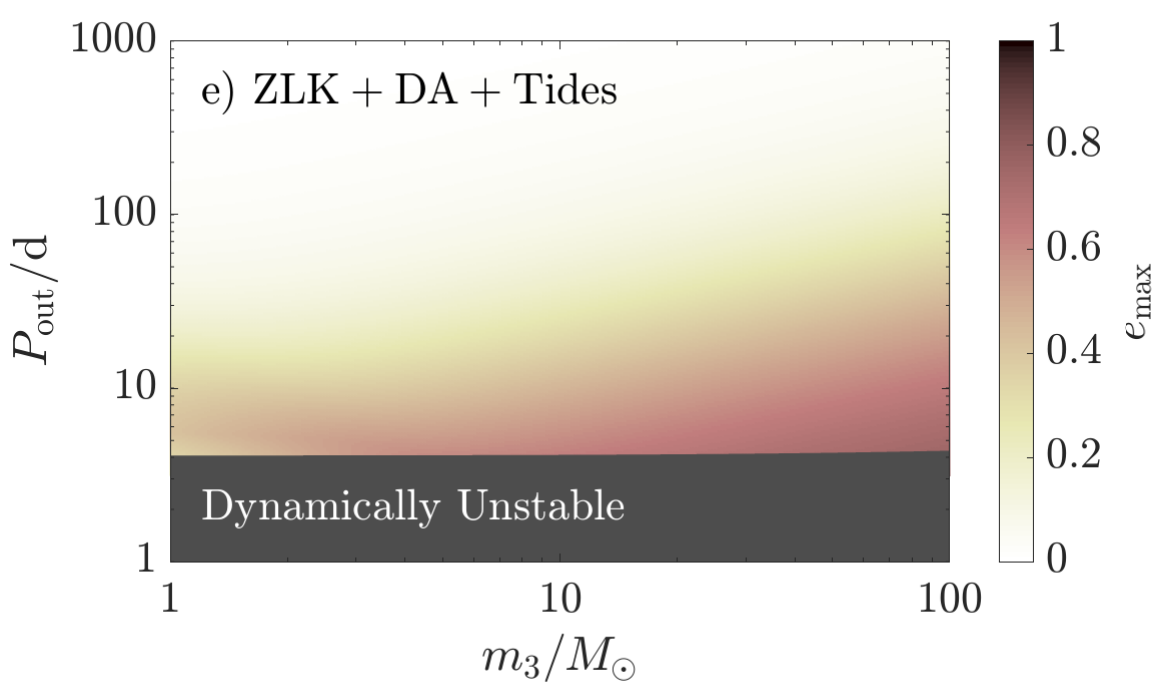}    
    \includegraphics[trim={0.5cm 8.5cm 0.9cm 9.5cm},clip,width=\columnwidth]{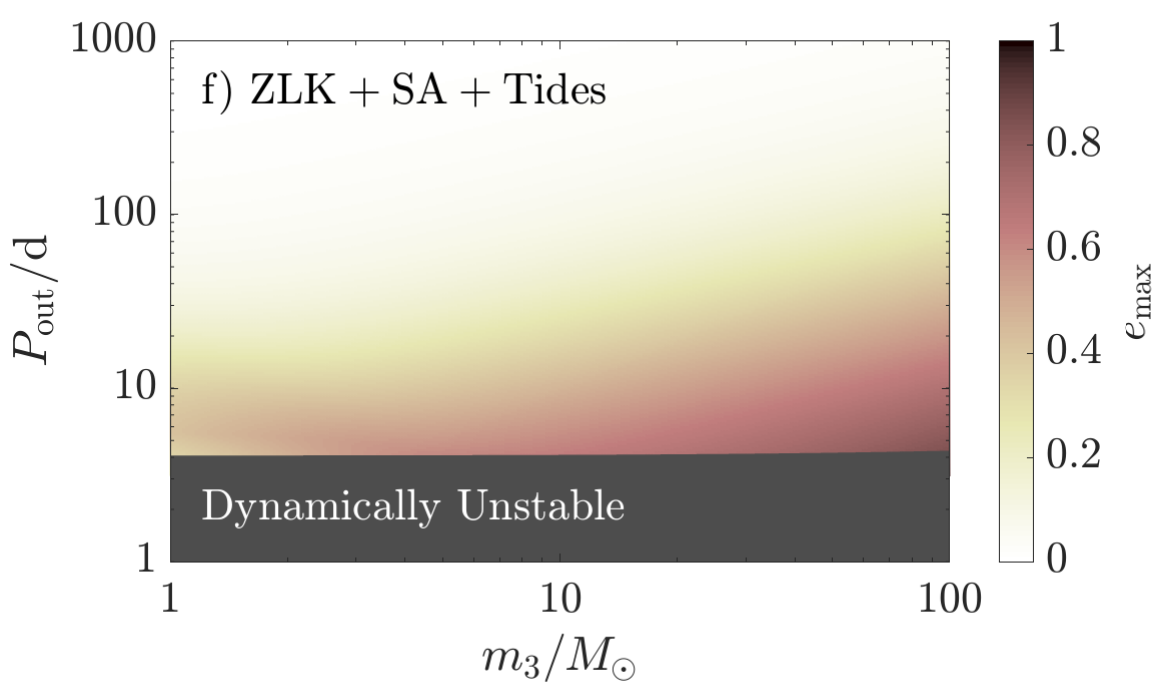}
    \includegraphics[trim={0.7cm 8.5cm 0.7cm 9.5cm},clip,width=\columnwidth]{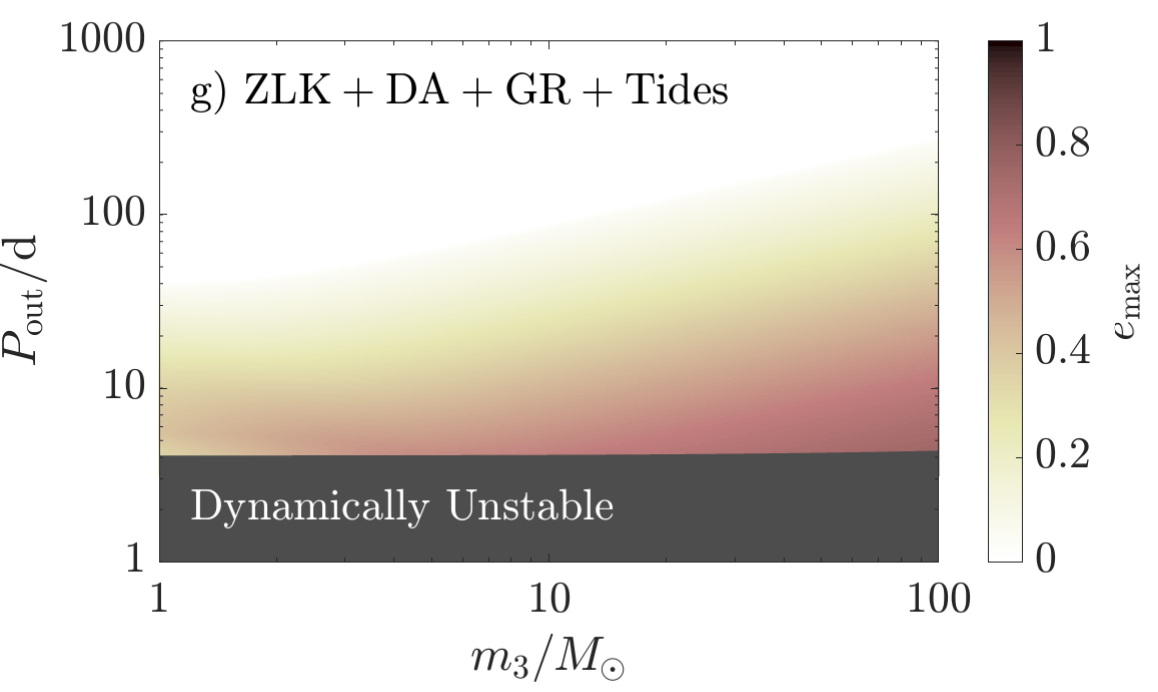}
    \includegraphics[trim={0.5cm 8.5cm 0.9cm 9.5cm},clip,width=\columnwidth]{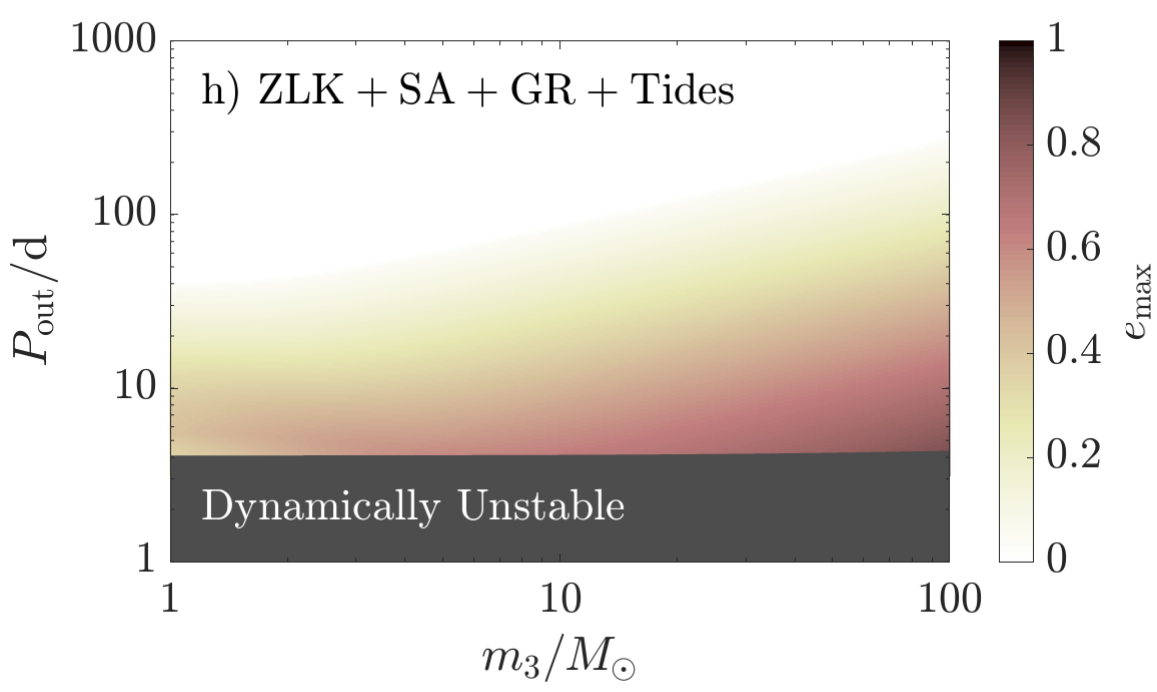} 
    \caption{
    Maximal inner eccentricity ($e_{\rm max}$) dynamically induced, via the ZLK mechanism, by a tertiary for a triple system with an arbitrary tertiary mass ($m_3$) and outer orbital period ($P_{\rm out}$). 
    The system includes an inner circular binary with masses $m_1=m_2=55\ M_{\odot}$ and an orbital period of 1.1 d, at a metallicity $Z=0.00042$, evaluated at the ZAMS.
    We use different colours to highlight various regions of interest.
    Dark grey indicates dynamically unstable triples.
    If the tertiary is unable to torque the inner binary, the latter remains circular ($e_{\rm max}\approx 0$); white regions indicate this regime.   
    We explore the impact of different assumptions (Sect. \ref{sec:meth:analytics}).
    The left and right columns show the results using the double-averaged (DA) and single-averaged (SA) approximations, respectively. 
    From top to bottom, the models incorporate additional terms that account for relativistic apsidal precession (GR), apsidal precession from the tidal equilibrium bulge in the inner binary (tides), and both.
    }
    \label{fig:short-range-analysis}
\end{figure*}

For our default CHE binary (Sect. \ref{sec:meth}), we ran a grid in $m_3$, $a_{\rm out}$, and $i_0$.
For $m_3$ and $a_{\rm out}$, we created a 250$\times$250 $\log$-uniform grid in the range $1\leq m_3/M_{\odot} < 100$ and $2a_{\rm in}\leq a_{\rm out} < 200a_{\rm in}$, which covers the parameter-space of interest between $1 < P_{\rm out}/\rm{d} < 1000$.
For $i_0$, we created a uniform grid with 181 grid points between $\cos(i_0)=\{-1, 1\}$, making sure to include $\cos(i_0)=0$.
We numerically solved Eq.~\eqref{eq:emax} for $e_{\rm max}$ using the bisection method, which then allowed us to calculate the fraction of stellar mergers (see Sect. \ref{sec:meth:merger_criteria} for our criteria).

Here, we present the results of our analysis focusing on the following aspects: the maximum eccentricity under various combinations of short-range forces (Sect. \ref{sec:res:maximum_eccentricity}), the predicted fraction of stellar mergers (Sect. \ref{sec:res:fraction_of_mergers}), and the broader implications for populations of stellar binaries and BBHs (Sect. \ref{sec:res:effects_on_populations}).

\subsection{Maximum eccentricity}\label{sec:res:maximum_eccentricity}
To assess the role of a tertiary companion in our CHE binary at the ZAMS, we first focus on the maximum eccentricity that can be induced via the ZLK mechanism. Figure \ref{fig:short-range-analysis} shows the maximum eccentricity for 8 different models (see Sect. \ref{sec:meth:triple_dynamics} for details).
The left column of Fig. \ref{fig:short-range-analysis} shows the DA approximation, which neglects perturbations occurring on timescales shorter than the outer orbital period \citep[e.g.][]{2022ApJ...934...44M}, while the right column includes models with the SA approximation. 
From top to bottom, the models account for short-range forces, as described below.
\begin{itemize}
    \item Panel (a) shows the result using exclusively the DA approximation.
    For tertiary companions with masses of $m_3 \gtrsim 10 M_{\odot}$, the outer orbital angular momentum dominates ($\eta \ll 1$, Fig. \ref{fig:adim_quantities}) and we are close to the test-particle limit, where the maximum eccentricity ($e_{\rm max} \approx 1$) is achieved for perpendicular orbits ($i_0=90$ deg).
    For less massive tertiaries ($m_3 \lesssim 10 M_{\odot}$), the inner angular momentum of the massive CHE binary can become dominant ($\eta > 1$), leading to a departure from the test-particle limit and damping the maximum eccentricity.
    The degeneracy on $e_{\rm max}$ in the $m_3$--$P_{\rm out}$ parameter space is inherited from the degeneracy on $\eta$ (Fig.~\ref{fig:adim_quantities}).
    \item Panel (b) presents the results obtained by relaxing the DA approximation to the SA approximation. 
    For our CHE binary, both approximations produce virtually identical results.
    \item Panel (c) presents the results obtained when accounting for the relativistic apsidal precession \citep{2001ApJ...562.1012E,2015MNRAS.447..747L}. 
    For tertiaries in outer orbital periods $P_{\rm orb}\gtrsim 100$ d, corresponding to $\epsilon_{\rm GR} \gtrsim 1$ (Fig. \ref{fig:short_range_forces}), relativistic precession can suppress the ZLK mechanism.
    For less massive ($m_3 \lesssim 10 M_{\odot}$), tight triples ($P_{\rm out} \lesssim$ 50 d), relativistic precession becomes negligible ($\epsilon_{\rm GR} < 1$, with $\eta>1$ and $\epsilon_{\rm SA} > \epsilon_{\rm GR}$).
    \item Panel (d) displays the analysis considering both the SA approximation and GR effects. 
    The results are essentially identical to those in panel (c).    
    \item Panel (e) shows the results when accounting for the equilibrium tide effect on the inner binary, where tidal bulges raised on the stellar surfaces induce a quadrupole moment that leads to apsidal precession \cite[e.g.][and references therein]{2021MNRAS.502.4479H}.
    Tides efficiently suppress the ZLK mechanism, leading to only small eccentricities ($e_{\rm max} \lesssim 0.2$) when the outer orbital period is long ($P_{\rm out} \gtrsim 100$ d), specifically when $\epsilon_{\rm Tide} \gg 1$. 
    Even for the most massive ($m_3=100 M_{\odot}$) and closest ($P_{\rm out}\approx 5$ d) tertiaries in our parameter space, the combined effect of ZLK and tides only leads to moderately large eccentricities ($e_{\rm max} \lesssim 0.75$). 
    \item Panel (f) displays the analysis considering both the SA approximation and the effect of tides. 
    The results are essentially identical to those in panel (d).
    \item Panel (g) displays the analysis when considering both  GR and tidal effects.
    In this case, tight systems ($P_{\rm out} \lesssim$ 50-100 d) experience ZLK quenching due to tides, while wider systems ($P_{\rm out} \gtrsim$ 50~--~100 d) experience ZLK quenching due to GR.
    \item Panel (h) includes the SA approximation, as well as GR and tidal effects; henceforth, this will be our default model, unless stated otherwise.
    The results are essentially identical to those in panel (g).    
\end{itemize}

We investigated the impact of various assumptions to assess how each one influences the population. 
Our findings confirm that relaxing simplifying assumptions and incorporating the effects of short-range forces mitigates the ZLK effect. 
In the point-mass test-particle approximation, a triple with $i_0=90$ deg can lead to $e_{\rm max} \approx 1$ on the ZLK timescale.
In addition, our analysis highlights the influence of short-range forces.
For instance, in our default model (ZLK+SA+GR+Tides, panel (h)), a tertiary with $m_3=10 M_{\odot}$ can attain $e_{\rm max}\approx 0.5$ ($e_{\rm max}\approx 0.1$) when $P_{\rm orb} \approx 50$ d ($P_{\rm orb} \approx 10$ d).
This maximum eccentricity becomes significantly larger when tides are negligible (ZLK+SA+GR, panel (e)), as the same triple configuration can lead to $e_{\rm max}\approx 0.9$ ($e_{\rm max}\approx 0.99997$) when $P_{\rm orb} \approx 50$ d ($P_{\rm orb} \approx 10$ d).
This scenario might occur if an inner stellar binary were to become a BBH with the same orbital parameters.

\subsection{Fraction of stellar mergers}\label{sec:res:fraction_of_mergers}

In Sect. \ref{sec:res:maximum_eccentricity} we utilise our default CHE binary model at the ZAMS (Sect. \ref{sec:meth:stellar_models}) to determine the maximum eccentricity attained in a triple configuration with a tertiary mass in the range $1 \leq m_3/M_{\odot} \leq 100$ and outer orbital period $1 \leq P_{\rm {out}}/\rm{d} \leq 1000$.
Those results represent a single realisation within the parameter space of initial mutual inclinations, corresponding to the angle in our grid that results in the largest eccentricity.
In this section, we examine the whole range of inclinations to assess the fraction of stellar mergers (see Sect. \ref{sec:meth:merger_criteria} for our criteria).
For each point in the grid of $m_3$ and $a_{\rm out}$, we estimate a merger fraction ($f_{\rm merger}$).
This fraction quantifies the proportion of inclinations, relative to a uniform distribution in $\cos(i_0)$, that result in mergers.

\begin{figure*}
    \centering
    % [trim={left bottom right top},clip] 
    \includegraphics[trim={0.7cm 8.5cm 0.7cm 9cm},clip,width=\columnwidth]{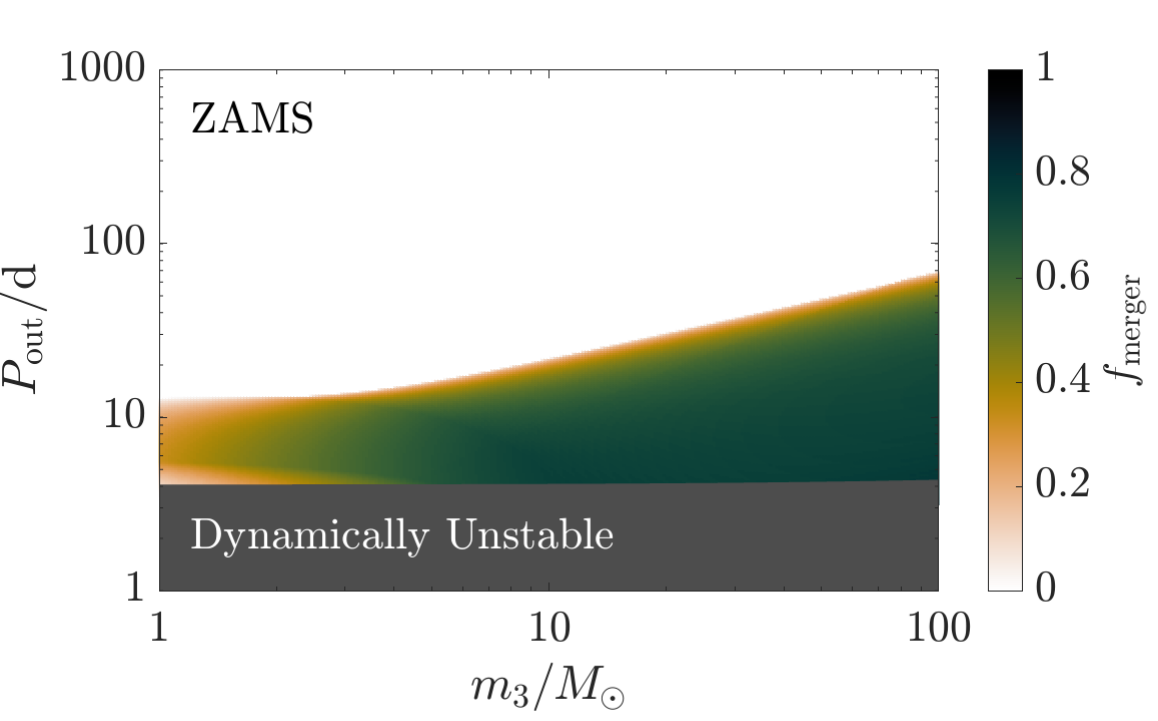}
    \includegraphics[trim={0.5cm 8.5cm 0.69cm 9cm},clip,width=\columnwidth]{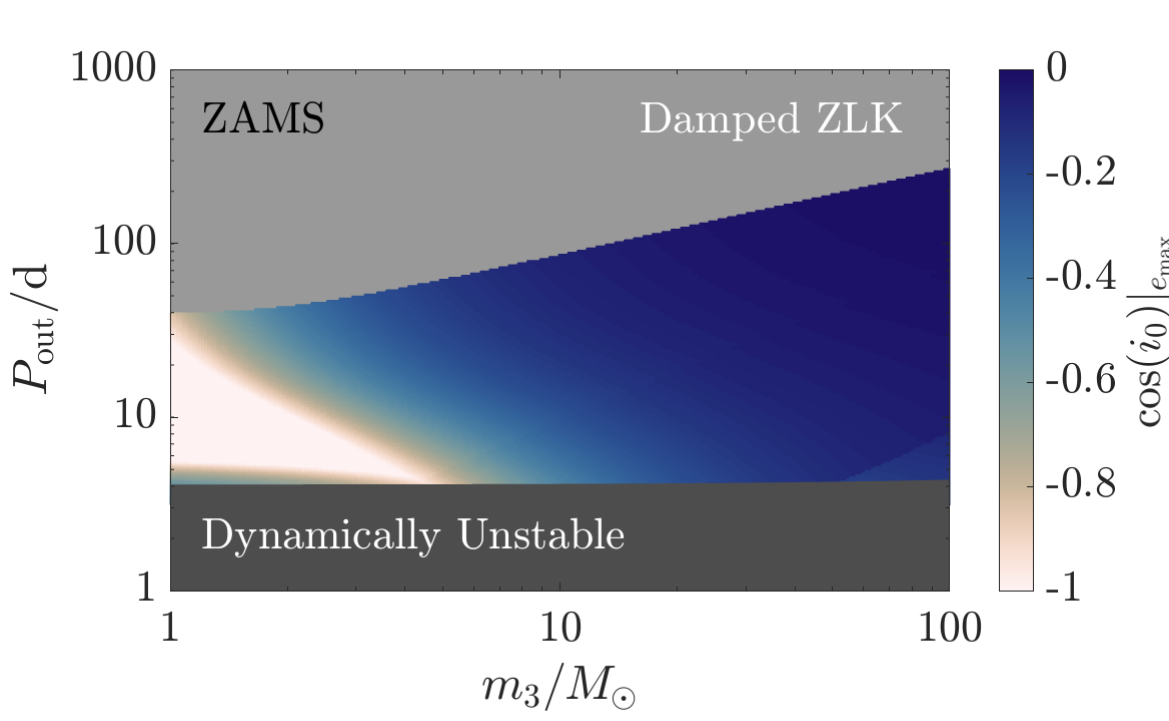}
    \includegraphics[trim={0.7cm 8.5cm 0.7cm 9cm},clip,width=\columnwidth]{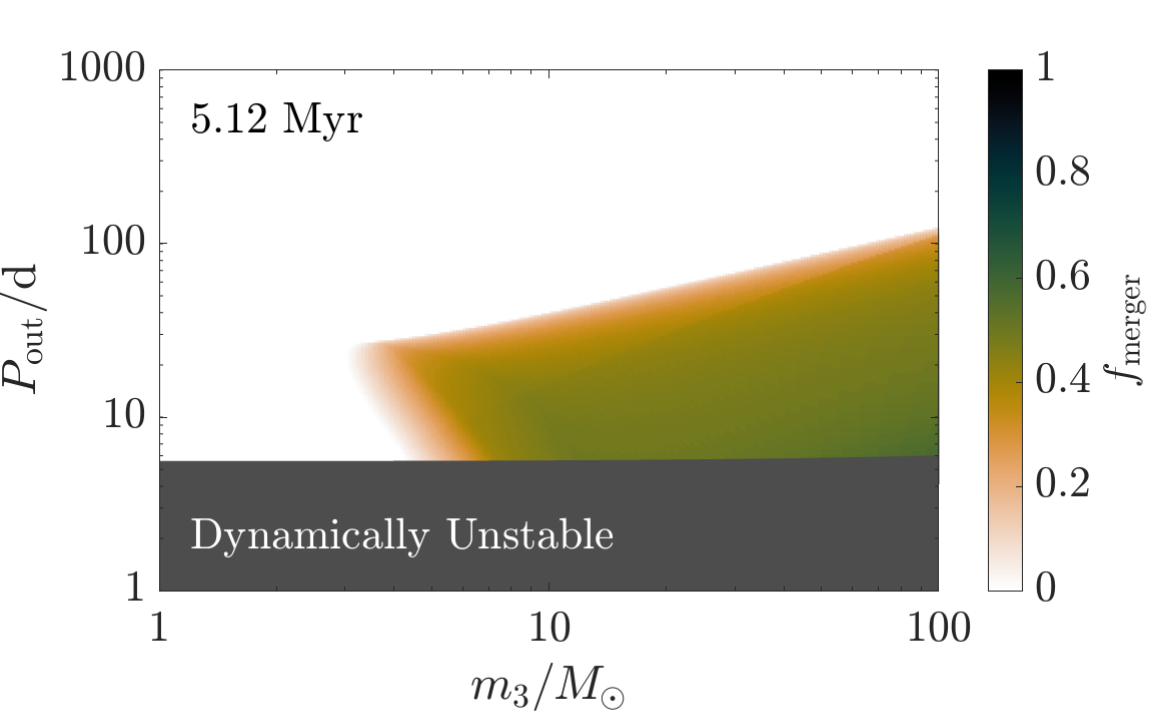}
    \includegraphics[trim={0.5cm 8.5cm 0.69cm 9cm},clip,width=\columnwidth]{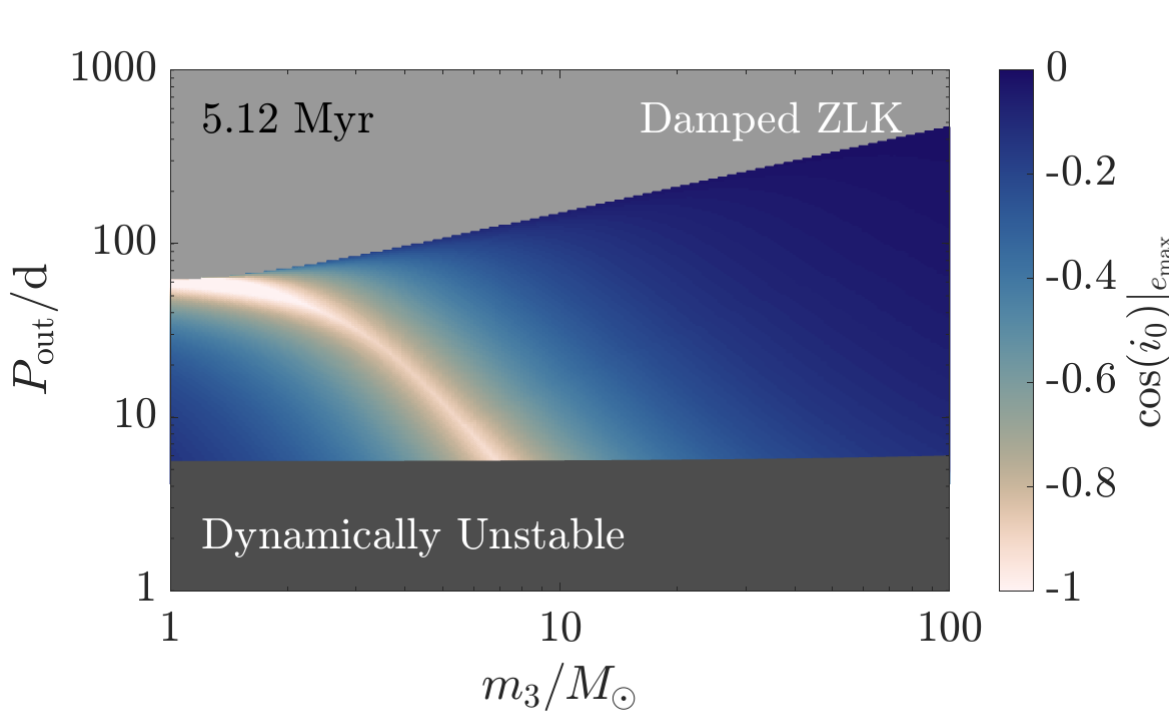}
    \caption{Merger analysis for a triple system with an arbitrary tertiary mass ($m_3$) and outer orbital period ($P_{\rm out}$), specifically addressing the ZLK mechanism including the single-averaged approximation, as well as the role of apsidal precession from GR and tides (Sect. \ref{sec:meth:analytics}).
    The system includes an inner circular binary with masses $m_1=m_2=55\ M_{\odot}$ and an orbital period of 1.1 d, at a metallicity $Z=0.00042$.
    We use different colours to highlight various regions of interest.
    Dark grey indicates dynamically unstable triples.    
    Specifically, we show the merger fraction ($f_{\rm merger}$, left panels), derived from an uniform grid in initial mutual inclinations inclinations ($\cos(i_0)=\{-1, 1\}$), as well as the value of $\cos(i_0)$ that results in the maximum eccentricity (right panels).
    We display the analysis for our chemically homogenously evolving binary at the ZAMS (top panels) and at $5.12$ Myr (bottom panels), shortly after the end of the main sequence (cf. Fig. \ref{fig:stellar_structure_low_Z}).
    In the right panels, grey indicates the region where the torque induced by the tertiary on the inner binary is not able to overcome the effect of tides.
    }
    \label{fig:f_merger_CHE}
\end{figure*}

Figure \ref{fig:f_merger_CHE} displays $f_{\rm merger}$ and the initial mutual inclination that lead to the maximum eccentricity ($\cos(i_0)|_{e_{\rm max}}$) for our default CHE binary. 
We explore these quantities at two different evolutionary stages: at the ZAMS and shortly after the end of the main sequence. For the latter, we consider the structure of our CHE model at 5.12 Myr, when the stellar model has contracted and initiated core-helium burning (Fig. \ref{fig:stellar_structure_low_Z}).

In the top-left panel of Fig. \ref{fig:f_merger_CHE}, we present $f_{\rm merger}$ near ZAMS. 
The $f_{\rm merger}$ fraction traces the $e_{\rm max}$ values as shown in panel (h) from Fig. \ref{fig:short-range-analysis}. 
This similarity arises because, for any fixed inner binary, there is a threshold eccentricity that satisfies our merger criteria. 
In this case, this eccentricity threshold is $e\approx 0.3$, resulting in a sharp distinction between triple configurations that lead to a merger, with $f_{\rm merger} \gtrsim 0$, and those that do not, with $f_{\rm merger} \approx 0$. 
For the most massive ($m_3=100 M_{\odot}$) and closest ($P_{\rm out}\approx 5$ d) tertiaries in our parameter space, the merger fraction is approximately $f_{\rm merger} \approx 0.77$.
The value of $f_{\rm merger}$ corresponds to an initial window in inclinations leading to $e \gtrsim 0.3$; at the ZAMS, this condition satisfies our merger criteria ($R>0.5r_{\rm p}$).
    
In the top-right panel of Fig. \ref{fig:f_merger_CHE}, we present the value of $\cos(i_0)|_{e_{\rm max}}$ at the ZAMS.
We find that massive tertiaries in wide orbits have $\cos(i_0)|_{e_{\rm max}} \approx \cos(90\ \rm{deg})=0$, which confirms the results in the point-mass test-particle approximation.
However, when the test-particle approximation is relaxed, less massive tertiaries in tight orbits deviate from this value, and the maximum eccentricity is achieved in highly-inclined retrograde orbits. 
For low-mass ($m_3 \lesssim 5 M_{\odot}$) close ($P_{\rm out} \lesssim 50$ d) tertiaries, there exists a significant portion of the parameter space where retrograde orbits, characterised by $\cos(i_0)|_{e_{\rm max}}=\cos(180\ \rm{deg})=-1$, lead to the maximum eccentricity.
The transition from $\cos(i_0)|_{e_{\rm max}}=0$ to $\cos(i_0)|_{e_{\rm max}}=-1$ signifies not only a departure from the test-particle approximation ($\eta > 1$), but also a regime where short-range forces start to play a role ($\epsilon_{\rm GR,Tide}>1$).
In this regime, the `window' of inclinations becomes narrower and more asymmetric around $\cos(i_0)|_{e_{\rm max}}$ \citep[cf. Fig. 1 in][]{2017MNRAS.467.3066A}, resulting in a decreased number of mergers.

In the bottom-left panel of Fig. \ref{fig:f_merger_CHE}, we display $f_{\rm merger}$ shortly after the end of the main sequence.
At that stage, the main change in the structure of the stars of the inner binary arises from the contraction of the helium-rich stellar components to $\approx 2.4 R_{\odot}$. 
This contraction has two competing effects. 
First, as presented in Eq. (\ref{eq:epsilon_tide}), it reduces the relative strength of tides.
Second, the threshold eccentricity required for a merger to occur has increased to $e \approx 0.8$. 
This change is also due to the orbital widening caused by wind mass loss, which leads to $P_{\rm in}=1.5$ d at this stage ($P_{\rm in}=1.1$ d at the ZAMS).
In massive wide triples, where $\eta \ll 1$, the parameter space that leads to mergers widens slightly.
In low-mass tight triples, where $\eta \gg 1$, tides and GR are not as dominant as they are at the ZAMS, and the quenching of ZLK comes from relaxing the test-particle limit.
Overall, the fraction of mergers is smaller than at the ZAMS.
For the most massive ($m_3=100 M_{\odot}$) and closest ($P_{\rm out}\approx 5$ d) tertiaries in our parameter space, the helium-rich merger fraction is approximately $f_{\rm merger} \approx 0.57$.
     
In the bottom-right panel of Fig. \ref{fig:f_merger_CHE}, we present the value of $\cos(i_0)|_{e_{\rm max}}$ shortly after the end of the main sequence.
For systems with tertiary masses $m_3 \gtrsim 10 M_{\odot}$, the results are very similar to those at the ZAMS (top-right panel).
We find that massive tertiaries in wide orbits have $\cos(i_0)|_{e_{\rm max}} \approx \cos(90\ \rm{deg})=0$.
Similarly to ZAMS, as we decrease tertiary mass we also reach a region where $\cos(i_0)|_{e_{\rm max}}=\cos(180\ \rm{deg})=-1$.
However, in contrast to ZAMS, for low-mass ($m_3 \lesssim 5 M_{\odot}$) close tertiaries ($P_{\rm out} \lesssim 50$ d), $\cos(i_0)|_{e_{\rm max}} \approx \cos(90\ \rm{deg})=0$.
This is because the effect of GR and tides is weak in that regime, leading back to the canonical solution for the ZLK effect \citep[cf. Fig. 1 in][]{2017MNRAS.467.3066A}.

These results underscore how stellar structure and binary evolution influence the dynamics of massive triple systems.
We specifically highlight two regimes that likely occur promptly around the ZAMS or after the end of the main sequence, resulting in hydrogen-rich and helium-rich stellar mergers, respectively.
We also show how the inclination that leads to the maximum eccentricity, along with the range of inclinations resulting in a threshold eccentricity required for merger, depends on the tertiary.
For tertiary objects with masses $m_3 \gtrsim 10 M_{\odot}$, where the outer orbital angular momentum dominates ($\eta \ll 1$; Fig. \ref{fig:adim_quantities}) close to the test-particle limit, the inclination that leads to the maximum eccentricity is 90 deg. 
Conversely, for tertiary objects with $m_3 < 10 M_{\odot}$, the inclination that results in maximum eccentricity is more sensitive to the stellar structure and may even be achieved in coplanar retrograde orbits.

\begin{figure}
    \centering
    % [trim={left bottom right top},clip] 
    \includegraphics[trim={0.7cm 8.75cm 0.7cm 9.5cm},clip,    width=\columnwidth]{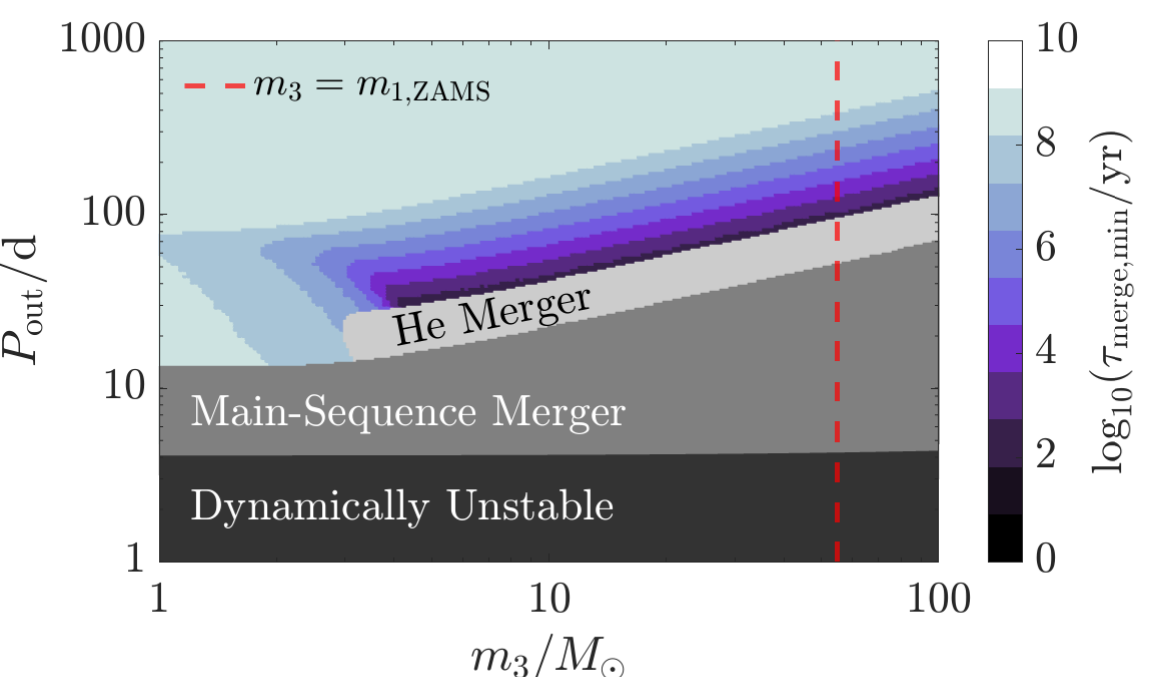}    
    \caption{Different outcomes in the parameter space of tight triples, with arbitrary tertiary mass ($m_3$) and outer orbital period ($P_{\rm out}$), that host inner binaries whose component stars undergo CHE.
    At the ZAMS, the system includes an inner circular binary with masses $m_1=m_2=55\ M_{\odot}$ and an orbital period of 1.1 d, at a metallicity of $Z=0.00042$.
    We use different colours to highlight various regions of interest.
    Dark grey indicates dynamically unstable triples.
    Grey and light grey represents the region where von Zeipel-Kozai-Lidov (ZLK) oscillations can lead to prompt hydrogen-rich and helium-rich stellar mergers in the inner binary, respectively.
    The colour bar gradient illustrates how ZLK oscillations can decrease the time-to-coalescence via gravitational-wave emission ($\tau_{\rm{merge,min}}$), potentially reducing it to the ZLK timescale (cf. Fig. \ref{fig:timescales}).
    Wide triples ($P_{\rm{out}}\gtrsim 1000$ d) remain unperturbed, with the inner binary evolving into a BBH that can merge in approximately $450$ Myr (Sect. \ref{sec:meth:stellar_models}). 
    The vertical dashed red  line marks the mass where $m_3=m_{1,\rm{ZAMS}}$, roughly delineating systems where the tertiary may evolve more rapidly than either of the inner binary components, potentially leading to a tertiary mass-transfer episode.
    }
    \label{fig:summary_low_Z}
\end{figure}

\subsection{Effects on populations}\label{sec:res:effects_on_populations}
After assessing the triple configurations that lead to stellar mergers (Sect. \ref{sec:res:fraction_of_mergers}), in this section we examine the impact of a tertiary companion at various evolutionary stages of our standard CHE binary model (Sect. \ref{sec:meth:stellar_models}). Figure \ref{fig:summary_low_Z} shows the summary of our analysis. 
We use different colours to highlight various regions of interest. 
The dark grey region indicates dynamically unstable triples.
The grey region delineates the parameter space where a tertiary object can induce a hydrogen-rich merger between the inner stars, provided that the mutual inclination promotes the ZLK mechanism.
This hydrogen-rich merger occurs near the ZAMS and on a timescale comparable to the ZLK timescale (Fig. \ref{fig:timescales}).
Similarly, the light grey region indicates the parameter space where an inner binary, which does not merge during the main sequence, can merge shortly afterwards, resulting in a helium-rich stellar merger.
This helium-rich merger would take place after the main sequence, either during the star's rapid contraction phase or after this contraction has ceased.
The parameter space of these mergers is narrower compared to hydrogen-rich mergers, yet it can lead to an intriguing merger remnant (see Sect. \ref{disc:stellar_mergers} for further discussion).

If a stellar merger is avoided, our default CHE binary can form a BBH.
For simplicity, we consider the case in which a BBH forms with identical parameters to those at the end of the stellar model simulation (Sect. \ref{sec:meth:stellar_models} and Fig. \ref{fig:stellar_structure_low_Z}).
Also in Fig. \ref{fig:summary_low_Z}, we denote in colour the minimum time to coalescence via gravitational-wave emission ($\tau_{\rm merge,min}$) for the inner BBH. 
If the BBH evolves in isolation or unperturbed by the tertiary, the time-to-coalescence is $\tau_{\rm merge} \approx 450$ Myr.
This is the case for triples with $P_{\rm out}> 500$ d.
However, there is a region of interest, dependant on tertiary mass, between $15 \lesssim P_{\rm out}/\rm{d} \lesssim 500$.
In this region, our default CHE binary might avoid a stellar merger and become a BBH.
Once the BBH forms, tidal forces are no longer present (i.e. $\epsilon_{\rm Tide}=0$). 
This absence allows the maximum induced eccentricity via the ZLK mechanism to increase, now primarily quenched through GR effects.
This induced eccentricity can lead to coalescence through gravitational-wave emission.
As demonstrated by \cite{2018ApJ...863...68L}, the merger timescale of such BBH is given by
\begin{equation}
    \tau_{\rm merge} \simeq T_{\rm c}(1-e_{\rm max}^2)^3,
\end{equation}
where $T_{\rm c}$ corresponds to the merger timescale for an isolated circular binary \citep{1964PhRv..136.1224P}.
For the most compact configurations that prevent stellar mergers, the time-to-coalescence can decrease by up to six orders of magnitude, resulting in a merger time of approximately $100$ yr.

Finally, in Fig. \ref{fig:summary_low_Z} we indicate with a vertical dashed red line where $m_3=m_{1,\rm{ZAMS}}$.
In a coeval triple system, where all component stars are born at the same time, tertiaries with masses $m_3>m_{1,\rm{ZAMS}}$ will evolve more rapidly than the stars of the inner binary.
This may lead to a tertiary mass-transfer episode, potentially influencing the future evolution of the triple system and causing deviations from the scenarios described here \citep[e.g.][]{2014MNRAS.438.1909D}.

\begin{figure}
    \centering
    % [trim={left bottom right top},clip] 
    \includegraphics[trim={0.7cm 8.75cm 0.7cm 9.5cm},clip,width=\columnwidth]{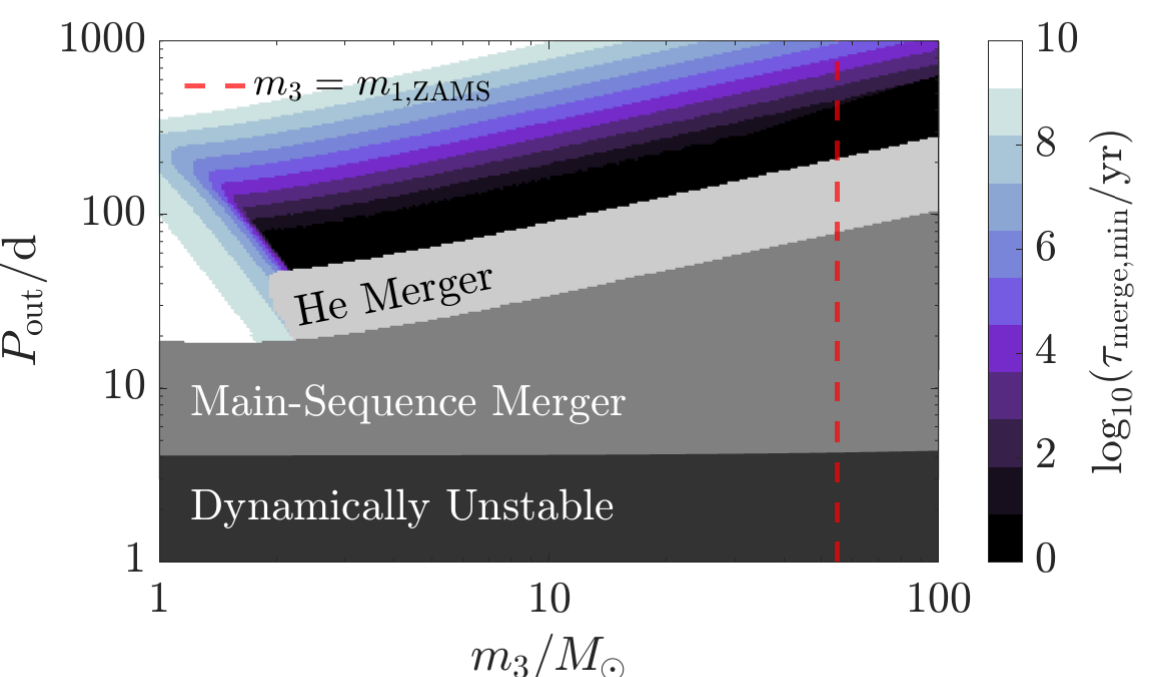}    
    \caption{
    Same as Fig. \ref{fig:summary_low_Z} but for $Z=0.0042$. Values of $\tau_{\rm merge,min} < 1$ yr are set to 1 year. The stellar model and merger analysis at this metallicity are provided in Appendix \ref{sec:appendix}.
    }
    \label{fig:summary_high_Z}
\end{figure}

Figure \ref{fig:summary_high_Z} shows the same exploration as in Fig. \ref{fig:summary_low_Z} but for a stellar model at $Z=0.0042$, which is twice as high as the commonly adopted value for the Small Magellanic Cloud \citep[SMC;][]{2024ApJ...966....9S}.
Except for the initial metallicity, which is ten times larger than in our default CHE (Sect. \ref{sec:meth:stellar_models}), all other initial parameters are identical (see Appendix \ref{sec:appendix} for details).
At this higher metallicity, the main difference is the increased mass loss due to stellar winds, which remove a combined total of $\approx 57\ M_{\odot}$($\approx 16\ M_{\odot}$) during(after) the main sequence and expand the orbit by a factor of a few (Appendix \ref{sec:appendix}).
This increased mass loss and orbital widening can lead to the formation of a BBH that would not merge within the age of the Universe. 
However, in the (tertiary mass-dependant) region where $P_{\rm out} \gtrsim 20$ d, there are many configurations where the ZLK mechanism can assist a merger within the age of the Universe, on timescales as short as the ZLK timescale.
Overall, we find that ZLK mechanism in CHE binaries can result in prompt stellar or BBH mergers.

%-----------------------------------------------------------------

\section{Discussion}
\label{sec:discussion}

\subsection{Mergers}
\subsubsection{Stellar mergers}\label{disc:stellar_mergers}

One of our goals in this paper is to establish a concise framework in order to investigate which orbital configurations lead to a stellar merger of the inner binary. 
These dynamical mergers occur rapidly \citep{2022MNRAS.515L..50V}, on timescales similar to the ZLK timescale, and reduce the triple system to a binary. 
Theory \citep[e.g.][]{2019Natur.574..211S} and observations \citep[e.g.][]{2024Sci...384..214F} tentatively suggest that merger remnants of massive stars are highly magnetic, at least immediately after the merger, with additional predictions for slow rotation \citep[e.g.][]{2019Natur.574..211S,2022NatAs...6..480W,2022MNRAS.517.2028K} and unusual abundances due to enhanced mixing \citep[e.g.][]{2005A&A...440.1041M}. 
Most merger scenarios discussed in the literature involve collisions \citep[e.g.][]{1976ApL....17...87H} or are initiated in isolated binary systems due to the radial expansion of at least one component star \citep[e.g.][]{1992ApJ...391..246P}, even when they occur within triples \citep{2022PhRvD.106b3014S} or quadruples \citep{2022MNRAS.515L..50V}. 
Here, we explore a scenario in which triple dynamics can aid a stellar merger during two specific phases in the evolution of stars.

First, we focused on hydrogen-rich mergers near the ZAMS. 
These occur once the inner binary has formed, shortly after the component stars have begun nuclear burning. 
These mergers are hydrogen-rich, and their structure and remnant evolution may resemble those of stellar mergers occurring in pre-main-sequence and star-forming regions. 
In these systems, the primary outcome is that the merger remnant is unlikely to evolve into a CHE star. 
This is because angular momentum can be removed during the merger or redistributed shortly afterwards, and tides are no longer present to maintain rotation. 
This outcome could, at least partially, help explain why there are not yet any unambiguous CHE binary candidates.

Second, we examined helium-rich mergers after the main-sequence. 
This second phase occurs when CHE stars contract on a thermal timescale (Fig. \ref{fig:stellar_structure_low_Z}).
This contraction reduces the influence of tides, thereby diminishing the suppression of the ZLK mechanism. 
As a result, some binaries that did not merge during the main sequence may do so shortly thereafter.
These helium-rich mergers were first identified in \cite{2024MNRAS.527.9782D} through a population analysis. 
Although the parameter space leading to helium-rich mergers is more restricted compared to hydrogen-rich mergers (Fig. \ref{fig:summary_low_Z} and \ref{fig:summary_high_Z}), helium-rich mergers can produce unique remnants.
In principle, they could result in massive highly magnetised helium stars with $\lesssim 10^5$ yr remaining before stellar collapse.
Such stellar structures have been previously linked to long-duration gamma-ray bursts and Type Ic superluminous supernovae \citep{2006A&A...460..199Y,2018ApJ...858..115A}. 
Depending on the mass, they could also lead to (pulsational) pair-instability supernovae \citep[e.g.][]{2003ApJ...591..288H}.

Finally, we highlighted the region within our parameter space that is likely to be influenced by the tertiary, assuming it is a star.
This region is shown in Fig. \ref{fig:summary_low_Z} and \ref{fig:summary_high_Z} as a vertical dashed red line, indicating where $m_3=m_{1, \rm{ZAMS}}$.
In coeval star systems, more massive stars evolve more rapidly than their less massive counterparts, causing them to leave the main sequence first.
Canonical stellar models for non-rotating stars show rapid expansion shortly after the main sequence. 
In our tight triple system, such expansion could lead to tertiary tides \citep{2023MNRAS.521.2114G} and eventually result in a mass transfer episode from the tertiary to the inner binary. 
The evolution and outcome of this tertiary mass transfer episode are quite uncertain \citep{2014MNRAS.438.1909D,2020MNRAS.491..495D,2020MNRAS.493.1855D,2020MNRAS.498.2957C,2021MNRAS.507.2659G,2022ApJS..259...25H}. 
One possible outcome is a stellar merger of the inner binary \citep{2024MNRAS.527.9782D}, or the formation of very tight systems following a triple common-envelope episode \citep{2021MNRAS.500.1921G,2021ApJ...907L..19V}. 
In any case, mass transfer from the tertiary will envelop the inner binary in a gaseous medium \citep[e.g.][]{2019ApJ...884...22A}, a configuration likely to affect the three-body dynamics \citep[see also][for other recent investigations on the dynamics of intermediate-mass and high-mass triples, respectively]{2025ApJ...978...47S,2025arXiv250500071S}. 

\subsubsection{Binary black holes}
Our default CHE binary evolves effectively undisturbed if the tertiary is $m_3<55\ M_{\odot}$ and the outer orbital period is $P_{\rm out}\gtrsim 100$ d (Fig. \ref{fig:summary_low_Z}).
In this regime, the inner CHE binary does not merge, but leads to the formation of a BBH. 
Once a BBH is formed, tidal forces are no longer present and the GR precession becomes the only effect suppressing the ZLK mechanism.
This may result in triples where the ZLK effect can significantly perturb the eccentricity of the BBH. 
In our default CHE binary, in certain triple configurations, the time-to-coalescence via gravitational-wave emission can decrease by several orders of magnitude, leading to prompt mergers only $\sim100$ yr after BBH formation (Fig. \ref{fig:summary_low_Z}).
This decrease is even more prominent in the alternative CHE binary at higher $Z=0.0042$.
At that metallicity, the BBH would not merge within the age of the Universe unless dynamically assisted by a tertiary (Sect. \ref{sec:res:effects_on_populations}), and its merger timescale can be of the order of the ZLK timescale (Fig. \ref{fig:summary_high_Z}).

The effects of a tertiary companion around a BBH might be varied and are uncertain.
\cite{2024MNRAS.527.9782D} demonstrated that triples with CHE binaries can experience tertiary mass transfer onto a BBH, which according to \cite{2023MNRAS.522.1686M} can result in mass growth via hypercritical accretion.
Subsequently, \cite{2025A&A...693A..84K} examined these configurations in greater detail, discovering that gas accretion onto the inner BBH can facilitate drag-assisted mergers, potentially resulting in an electromagnetic counterpart \citep[see also][for an EM counterpart to a BBH collision with a massive star]{2025ApJ...983L...9K}.
In addition, here we find that some BBH mergers can promptly occur purely through dynamical assistance via the ZLK mechanism.
In close triples that lead to prompt BBH mergers, the dynamics may fall within a non-adiabatic regime \citep[e.g.][]{2018MNRAS.480L..58A,2018ApJ...863....7R,2020MNRAS.493.3920F}. 
In this regime, the black hole spins evolve too slowly to follow the rapid ZLK oscillations through relativistic spin-orbit coupling.
When a BBH merges, it losses some mass-energy through gravitational waves, and depending on the spin configuration, it may receive a recoil kick.
BBHs formed from CHE evolution are believed to have notably high spins, unlike the typically non-rotating black holes from other evolutionary channels in isolated binary evolution \citep{2024A&A...691A.339M}.
For isolated CHE binaries, these spins are thought to be aligned; however, it is not clear how black hole formation and precession might affect the direction of these spins \citep[e.g.][]{2022ApJ...938...66T}.
Depending on the alignment of these spins, the recoil kick for the black hole remnant could range from $\sim 10-1000$ km/s. 
Such a recoil kick could propel the black hole through surrounding material lost via winds or during black hole formation \citep[e.g.][]{2023ApJ...957...68B} during the late stages of evolution, as well as prompt interactions with the tertiary if it is a stellar companion. 
Therefore, we highlight these ZLK-assisted BBH mergers as potential candidates for joint gravitational-wave and electromagnetic detection. 

\subsection{Caveats}

\subsubsection{Stellar evolution}
Our analysis examines two stellar models with the same mass, but different metallicities. 
The evolution of massive stars is influenced by various uncertainties, including convection, mixing, mass loss, and magnetic fields. 
Here, we emphasise two main caveats: stellar expansion and collapse.

Stellar expansion following the main sequence is natural in most non-rotating models. 
In short-period binary systems, this expansion can lead to a mass transfer episode, possibly resulting in a merger during the main sequence.
According to \cite{2022MNRAS.515L..50V}, in systems where a merger is imminent, the dynamics will only cause the merger to occur sooner than in isolated systems.
This can result in BBH mergers where the stellar progenitors underwent a stellar merger \citep{2022PhRvD.106b3014S}.

In this paper, we focus on the CHE scenario, where rotational mixing allows the star to remain compact throughout the main sequence and leads to contraction afterwards. 
\cite{2021ApJ...907L..19V} noted that any compact star can play a significant role in the evolution of tight triples, including stars in metal-free (Population III) environments \citep[e.g.][]{2001A&A...371..152M} or within specific mass–metallicity regimes \citep[e.g.][]{2020A&A...634A..79S}. We adhere this idea and propose that prompt (stellar and BBH) mergers are likely to occur in systems where the component stars of the inner binary do not significantly expand after the main sequence, regardless of the physical mechanism behind this.

With respect to stellar collapse, we assume that our $\approx43.8\ M_{\odot}$ carbon-oxygen core will form a black hole \citep[e.g.][]{2003ApJ...591..288H,2012ApJ...749...91F}.
We further assume that this black hole is created through direct collapse, without any baryonic mass ejection or associated explosion. 
Energy is lost exclusively via neutrinos, resulting in a mass decrement of $\sim0.1\ M_{\odot}$ and no natal kick, consistent with some observed systems \citep[e.g.][and references therein]{2024PhRvL.132s1403V}, including triples \citep[e.g.][]{2024Natur.635..316B,2025ApJ...983..115S}.
However, it is important to acknowledge the possibility of mass ejection during black hole formation, as well as a natal kick. 
Mass ejection could contribute to the energy reservoir relevant for an electromagnetic counterpart to the BBH merger \citep[e.g.][]{2017ApJ...839L...7D,2023ApJ...957...68B}. 
A natal kick could affect the orbits and their dynamics. 
These effects are not considered in this study.

Finally, there is a more massive mass regime where an oxygen core undergoes pair-instability \citep{1967PhRvL..18..379B,1967ApJ...148..803R}. 
In this regime, efficient electron-positron pair production leads to the collapse of the oxygen core, raising the central temperature sufficiently to ignite explosive oxygen burning. 
If the explosion is strong enough, it can completely disrupt the star, leaving no remnant. 
If it is not as strong, the explosion may eject the outer layers of the star through one or more `pulsations', eventually resulting in the formation of a black hole. 
These pulsations could interact with stellar companions and affect the triple orbit. 
We do not account for these effects in our study.

\subsubsection{Binary evolution}

In our analysis, the role of tides in the inner binary is crucial (Sect. \ref{sec:meth:tides}).
Tides influence the evolution of binary and triple systems through various effects, such as rotation, tidal distortion, and tidal dissipation (Sect. \ref{sec:meth:tides:comp}).

Massive short-period binaries are often assumed to rotate synchronously and be tidally locked, a configuration believed to support and maintain rapid rotation throughout their stellar lifetimes \citep{2009A&A...497..243D}.
Rotation plays a critical role in tidally locked massive binary stars, as it drives meridional (pole-to-equator) circulation, known as Eddington–Sweet circulation \citep[][and references therein]{1986MNRAS.221...25M}. 
This circulation can induce mixing, leading to a homogeneous composition \citep{1987A&A...178..159M}, and can promote the formation of a CHE binary. 
Furthermore, rotation-induced oblateness can alter short-range potential terms relative to the quadrupole potential, potentially suppressing the ZLK mechanism \citep{2015MNRAS.447..747L}.

Besides rotation, the presence of a close companion can further tidally distort the star, enhancing mixing efficiency compared to models that consider only rotation \citep{2020A&A...641A..86H}.
Additionally, in a tidally deformed short-period binary a non-dissipative tidal bulge may form. 
This bulge can lead to apsidal precession, which suppresses the ZLK mechanism (Sect. \ref{sec:meth:analytics}).
Our default CHE binary is not only a rotating, short-period binary but also a contact system. 
In contact binaries, tidal distortion can cause a radius discrepancy of up to 5\% \citep{2022A&A...661A.123F}; however, this discrepancy does not significantly impact our calculations.

In our framework for estimating the maximum eccentricity (Sect. \ref{sec:meth:analytics}), we assume that tides influence the system through the potential created by the non-dissipative tidal bulge, with no energy dissipation occurring in the system.
However, in binaries with orbital periods less than approximately 10 days, tides serve as an efficient mechanism for energy dissipation. 
In this paper, we focused on the dynamical tide, which is suitable for stars with a radiative envelope, as the primary tidal dissipation mechanism (Sect. \ref{sec:meth:tides});  we estimated $\tau_{\rm{circ}} \approx 1500\ \rm yr$ and $\tau_{\rm{dyn}} \approx 23\ \rm yr$.
Within the formalism for the dynamical tide, there are major uncertainties in the value of the second-order tidal coefficient $E_2$ \citep[e.g.][]{2013A&A...550A.100S,2017MNRAS.467.2146K}, which is inversely proportional to $\tau_{\rm{circ}}$ (Eq. \eqref{eq:dynamical_tide_circ}).
Different models for of $E_2$ can lead to shorter ($\tau_{\rm{circ}} \approx 185$ yr and $\tau_{\rm{dyn}} \approx 3$ yr, \citealt{2018A&A...616A..28Q}) or longer ($\tau_{\rm{circ}} \approx 2400$ yr and $\tau_{\rm{dyn}} \approx 37$ yr, \citealt{2010ApJ...725..940Y}) estimates for the tidal timescale at the ZAMS and throughout the stellar lifetime.
These different parametrisations can either constrain or expand the parameter space where the ZLK mechanism influences the eccentricity of the inner binary.

Recently, \cite{2024A&A...681L...1S} addressed how the formulae from \cite{2002MNRAS.329..897H}, which are widely used to estimate the synchronisation timescales via the dynamical tide, are inconsistent with the original model from \cite{1977A&A....57..383Z}.
They propose an alternative formula to compute the circularisation timescale via the dynamical tide.
For nearly synchronous, equal-mass binaries, such as those in our Fiducial model, the synchronisation timescale should be considerably longer than the estimate provided in Sect. \ref{sec:meth:tides}. This implies that tidal dissipation in the inner binary becomes negligible (Sect. \ref{sec:meth:tides:comp}), unless the binary spins become significantly asynchronous due to perturbations by the tertiary.

Finally, we briefly discuss the role of wind mass loss, which becomes increasingly important at high metallicities (Fig. \ref{fig:stellar_structure_high_Z}). 
For the component stars, mass loss through winds ultimately influences the remnant mass of the black hole. 
More crucially, this mass loss alters the orbital separation --- a key factor in determining whether a BBH can merge within the age of the Universe \citep{1964PhRv..136.1224P}.
Our model assumes that each star loses mass via a fast isotropic wind that carries away the specific angular momentum of the mass-losing star. 
This mechanism, often termed the `Jeans mode' or `fast winds' mode \citep{1963ApJ...138..471H}, leads to orbital widening proportional to the amount of mass lost. Consequently, wide ($\sim 1000$ d) triples at moderately-low metallicity (Fig. \ref{fig:summary_high_Z}) do not result in BBH mergers, unlike similar initial configurations at low metallicity (Fig. \ref{fig:summary_low_Z}). 
While fast winds are appropriate for massive, close binaries, slow winds could induce drag and limit the orbital widening \citep[e.g.][]{2021arXiv210709675S}. 
Additionally, the orbital evolution of the inner binary might modify the triple dynamics, potentially triggering a dynamical instability or affecting ZLK cycles.

\subsubsection{Triple evolution}

In the study of multiple star systems, it is often assumed that the component stars are coeval. 
This assumption was applied in our analysis.
However, simulations indicate that some triples form after the inner binary has assembled, either as part of the same starburst \citep[e.g.][]{1994MNRAS.269L..45B} or through a dynamical capture.
In our tight triple systems, the ZLK timescale ($\tau_{\rm ZLK}\lesssim 10^4$ yr) is very short with respect to the stellar lifetime ($\sim 10^6$ yr).
Consequently, for a coeval triple system, a merger near the ZAMS would effectively correspond to a merger in the pre-main-sequence phase. 
Alternatively, the merger might occur at a much later stage, after the triple system has formed and achieved a dynamical steady-state equilibrium. 
Only then would it evolve in a manner more akin to the scenario considered in our formalism.
In such cases, the merger would not occur around the ZAMS but shortly after the triple assembled.

Once the triple system has formed, there are additional considerations to keep in mind. 
In our analysis, we have assumed that the mass ratio is close to one and that the outer orbit is circular.
This assumption for the mass ratio is common in the literature on CHE binaries \citep{2016A&A...588A..50M,2016MNRAS.458.2634M}. 
However, massive contact binaries are often observed with mass ratios that differ from unity \citep[see the sample presented in][]{2021MNRAS.507.5013M}.
While \citet{2023A&A...672A.175F} determined that energy transfer in massive contact binaries alters the evolution of the mass ratio and can prolong the lifespan of a contact system with unequal masses, \citet{2025A&A...695A.109F} showed that this effect is small at a population level.

The assumption about a circular outer orbit might not be representative for hierarchical triple systems \citep[e.g.][]{2022MNRAS.512.3383H}.
The two tight high-mass triples, TIC 470710327 \citep{2022MNRAS.511.4710E} and TIC 290061484 \citep{2024ApJ...974...25K}, exhibit outer eccentricities of $e_{\rm out}=0.3$ and $e_{\rm out}=0.2$, respectively.
Also, the tertiary around a massive binary may have an eccentric orbit due to the apsidal precession resonance \citep[e.g.][]{2024PhRvL.132w1403L}. 
For systems with non-equal mass binaries and non-zero outer eccentricity, the octupole effect may become important \citep{2016ARA&A..54..441N}. 
This tends to widen the inclination window for large eccentricity excitation.
However, the analytic expression for $e_{\rm max}$ given by Eq.~\eqref{eq:emax} remains valid. 
Thus, because of the effect of short-range forces due to tides, the maximum eccentricity cannot exceed $e_{\rm max}$ even when the octupole potential is significant \citep{2015MNRAS.447..747L}.

\subsection{Constraining tidal theory}

There are significant uncertainties in our understanding of tidal theory, particularly in massive, short-period binary systems. These uncertainties primarily stem from the standard model of the dynamical tide \citep{1975A&A....41..329Z,2008EAS....29...67Z}, which describes how stellar oscillation modes are excited by a periodic tidal potential. A key parameter in this model, the $E_2$ coefficient (Sect. \ref{sec:meth:tides}), is highly sensitive to the definition of the core-envelope boundary \citep[e.g.][]{2013A&A...550A.100S,2017MNRAS.467.2146K}, making it a crucial but uncertain factor in tidal evolution.

In tight triple systems, tidal effects introduce additional complexity. If the dynamical tide is strong enough, it may suppress the ZLK mechanism, preventing eccentricity oscillations in the inner binary. If suppression does not occur, additional tidal interactions could significantly alter the evolution of the inner binary. In eccentric systems, tidal dissipation differs from the standard case, as close periastron passages can induce oscillations that lead to `heartbeat'
stars \citep{2012ApJ...753...86T}.

The interplay between the inner and outer binaries remains highly uncertain. Short-range forces, including rotation, tides, and GR effects, have been shown to suppress the ZLK mechanism \citep{2015MNRAS.447..747L}. In our default CHE binary model, tides are the dominant short-range force responsible for this suppression. However, the structure and evolution of the tidal bulge, which determines the maximum eccentricity of the inner binary and whether it will merge as a stellar or BBH system, are still not well understood.

To refine tidal theory, we propose using future observations of high-mass tight triple systems. The mere detection of such systems could provide empirical constraints on tidal effects. If the orbital configuration suggests that the ZLK mechanism should be active, we can compare cases with and without tidal suppression (ZLK+SA+GR+Tides vs ZLK+SA+GR) to determine whether tides are strong enough to prevent eccentricity growth. If the inner binary remains circular over the ZLK timescale (Sect. \ref{sec:meth:tides}), this would place an upper limit on the tidal dissipation timescale. Additionally, such systems could allow us to solve for  $E_2$  in specific binary configurations, further improving tidal models. Conversely, if the inner binary exhibits eccentricity evolution, we can compare the observed evolution to our ZLK model to assess the accuracy of our tidal prescriptions, either by determining the value of $\epsilon_{\rm Tide}$ or by quantifying deviations from the expected ZLK+SA+GR solution.

In summary, observations of tight triple systems offer a direct way to test and refine tidal theory. By analysing their orbital configurations and evolutionary pathways, we can place new constraints on tidal dissipation, improve our understanding of how tides suppress eccentricity growth, and ultimately enhance our models of stellar and compact object mergers.

\subsection{Population forecasting for tight triples}
In a triple star system, a merger event reduces the system to a binary configuration. 
The characteristics of binaries that originate from triples remain an open question. 
As demonstrated here, certain triple configurations can lead to prompt mergers. 
For triple star systems, these prompt mergers likely have two effects: they result in a binary system with a massive merger remnant and suggest that some system configurations are unlikely to be observed as triples. 
In this study, we tried to remain as neutral as possible regarding the initial distribution of tight triples.
However, we find that stellar mergers depend on the initial properties of the tertiary mass, separation, and mutual inclination. 
Therefore, any population-level predictions will be sensitive to assumptions about the initial orbital distributions.
Moreover, merger hydrodynamics are very uncertain and not yet studied in the context of the ZLK mechanism.
Additionally, for binaries where mergers are imminent even without a tertiary companion, the ZLK mechanism can facilitate the merger process, causing it to occur more quickly or at an earlier evolutionary stage.

Finally, tight triples in which the inner binary forms a BBH present very interesting prospects.
These can result in triple systems with a tertiary in an outer orbital period of 100 or 1000 d, detectable with facilities such as Gaia \citep{2016A&A...595A...1G}.
Triple configurations that lead to a prompt BBH merger could potentially be detected by the LIGO-Virgo-Kagra detector network,  resulting in a binary configuration of an unusually massive black hole orbited by a nearby stellar companion.
%-------------------------------------- Two

%-----------------------------------------------------------------

\section{Summary and conclusions}
\label{sec:summary_and_conclusions}

We investigated the role of the ZLK mechanism in high-mass triple systems.
We focused on `tight' triples, which are composed of a short-period, high-mass stellar binary with a nearby tertiary companion. 
We explored an illustrative case involving a circular binary with component masses $m_1=m_2=55\ M_{\odot}$, initial orbital period $P_{\rm in}=1.1$ d at 'low' ($Z=0.00042$) and `moderately low' ($Z=0.0042$) metallicities (Sect. \ref{sec:meth:binary_evolution} and Appendix \ref{sec:appendix}, respectively).
This binary was simulated using the detailed stellar evolution software MESA.

The component stars in this binary undergo chemically homogeneous evolution (CHE), resulting in well-mixed stars that remain compact throughout the main sequence and contract afterwards (Fig. \ref{fig:stellar_structure_low_Z} and \ref{fig:stellar_structure_high_Z}). 
Based on our numerical model, we examined the impact of a tertiary with a mass range between $1 \leq m_3/M_{\odot} \leq 100$ and an outer orbital period range between $1 \leq P_{\rm out}/\rm{d} \leq 1000$, assuming the initial mutual inclination is uniformly distributed in the range of $-1 \leq \cos(i_0) \leq 1$.

We utilised the analytic results from \cite{2015MNRAS.447..747L} and \cite{2022ApJ...934...44M} to determine the maximum eccentricity that the inner binary can achieve in different triple configurations (Sect. \ref{sec:meth:triple_dynamics} and Fig. \ref{fig:short-range-analysis}). 
We evaluated whether the ZLK mechanism could cause the component stars of the binary to fill their outer Lagrangian points. 
If this occurs, we consider the inner binary to result in a stellar merger (Sect. \ref{sec:res:fraction_of_mergers}).
We find three key evolutionary stages for mergers: near the ZAMS, shortly after the end of the main sequence, or after the formation of a BBH (Figs. \ref{fig:summary_low_Z} and \ref{fig:summary_high_Z}).
 Our concluding remarks are as follows.
\paragraph{Main-sequence mergers.}
During the main sequence, strong tidal bulges in close binaries significantly damp the maximum eccentricity produced by the ZLK mechanism (Figs. \ref{fig:short_range_forces} and \ref{fig:short-range-analysis}). 
Nonetheless, mergers can occur even at moderately-low eccentricities ($e\gtrsim0.3$), with an occurrence fraction of $f_{\rm{merger}}\approx 0.77$ for optimal inclinations. 
These mergers are possible in systems with outer periods of up to $\approx 70$ d at low metallicity (Fig. \ref{fig:f_merger_CHE}) and up to $\approx 100$ d at moderately-low metallicity (Fig. \ref{fig:f_merger_CHE_high_Z}). 
Given the ZLK timescales (Fig. \ref{fig:timescales}), these hydrogen-rich mergers must happen near the ZAMS for coeval triples or shortly after the formation of the triple for non-coeval systems.
\paragraph{Post-main-sequence mergers.}
After the main sequence, CHE stars rapidly contract and the effect of tides weakens. 
At this stage, mergers can occur at high eccentricities ($e \gtrsim 0.8$), and occur less frequently at a fraction of $f_{\rm merger}\approx 0.57$ for optimal inclinations.
These mergers are helium-rich and can occur in systems with outer periods of up to $\approx 100$ d at low metallicity (Fig. \ref{fig:f_merger_CHE}) and up to $\approx 270$ d at moderately-low metallicity (Fig. \ref{fig:f_merger_CHE_high_Z}). 
\paragraph{Binary black hole mergers.}
For triples in which the inner binary does not undergo a stellar merger, a BBH may form. 
In BBHs, tides are no longer present, and the ZLK mechanism is largely suppressed by GR effects. 
The merger of BBHs can be facilitated by a tertiary companion with an orbital period of up to a few thousand days, and in favourable orbital configurations the time-to-coalescence may be as short as the ZLK timescale (Fig. \ref{fig:summary_low_Z} and \ref{fig:summary_high_Z}). 
This results in prompt mergers shortly after the formation of a BBH within a tight triple system, which contrasts with the generally slower coalescence expected for BBHs formed in isolation.
These BBH mergers occur near recently ejected gas (from stellar winds or following core collapse) and possibly close to another star. 
This makes them promising candidates for electromagnetic counterparts and probes of environmental effects in gravitational-wave sources.
\paragraph{Orbital configurations leading to mergers.}
Contrary to the test particle limit, maximum eccentricity is not always achieved at a $90$ deg mutual inclination. Instead, it depends on the ratio of the inner to outer orbital angular momenta ($\eta := L_{\rm in}/L_{\rm out}$, Fig. \ref{fig:adim_quantities}). 
Moreover, when the tertiary is less massive than the inner component stars of the binary, the maximum eccentricity is reached at largely retrograde configurations (Fig. \ref{fig:f_merger_CHE} and \ref{fig:f_merger_CHE_high_Z}).
This behaviour is independent of the nature of the component stars of the inner binary; it should also occur in stars that do not evolve via CHE.
\paragraph{Outlook.}
The dynamics of tight, high-mass stellar triples are complex. 
When the tertiary star is more massive than either component stars of the inner binary, the evolution of the system approximates the test particle limit. 
However, in this regime, stellar evolution (particularly mass transfer from the tertiary) can play a crucial role. 
Conversely, if the tertiary is less massive than either inner companions, evolution is dominated by the inner binary, and even significantly inclined systems may experience reduced dynamical effects due to being far from the test particle limit. 
On a population level, an important question remains as to which tight triples are eventually reduced to binaries and which remain intact as triples, noting that the latter stand as candidate progenitors of merging BBHs.

\begin{acknowledgements}
The authors thank Sunmyon Chon, Selma de Mink, Ilya Mandel, Luca Sciarini, Silvia Toonen, and Ruggero Valli for helpful discussions.
We thank the referee Ying Qin for useful suggestions.
EG and AB acknowledge support from the Australian Research Council (ARC) Centre of Excellence for Gravitational Wave Discovery (OzGrav), through project number CE230100016.
EG further acknowledges support from the ARC Discovery Program DP240103174 (PI: Heger).
BL acknowledges support from National Natural Science Foundation of China (Grant No. 12433008).
This research made use of MESA: Modules for Experiments in Stellar Astrophysics, r11701 
\citep[\url{http://mesa.sourceforge.net},][]{2011ApJS..192....3P,2013ApJS..208....4P,2015ApJS..220...15P,2018ApJS..234...34P,2019ApJS..243...10P}.
The MESA inlists used used to model binary evolution are available via zenodo, with DOI 10.5281/zenodo.3667546 \citep{l_du_buisson_2019_3667546}.
The scripts used for the dynamical analyses in this study are available on GitHub at \href{https://github.com/avigna/che-triples}{avigna/che-triples}.
Kippenhahn diagram plotter for MESA available in GitHub at \href{https://github.com/orlox/mkipp}{orlox/mkipp}.
\end{acknowledgements}

\bibliographystyle{aa} % style aa.bst
\bibliography{references} % your references Yourfile.bib

\begin{appendix}

\section{Analysis at $Z=0.0042$}\label{sec:appendix}
In this paper, we use detailed stellar evolution to explore the CHE of a binary (Sect. \ref{sec:meth:stellar_models}).
To do so, we followed the study of \cite{2024ApJ...966....9S},  focused on HD 5980, a multiple-star system located within the SMC. 
To model HD 5980, \cite{2024ApJ...966....9S} adopts a default metallicity of $Z = Z_{\odot}/4 = 0.00425$, using $Z_{\odot} = 0.017$ \citep{1996ASPC...99..117G}. 
However, they acknowledge that this value is twice as high as the commonly accepted metallicity for the  SMC.
Given that CHE is highly sensitive to metallicity \citep[e.g.][]{2017A&A...604A..55M,2021MNRAS.505..663R}, we explore an alternative model with higher metallicity while maintaining all other assumptions.

\begin{figure}[]
    \centering
\includegraphics[width=0.5\textwidth]{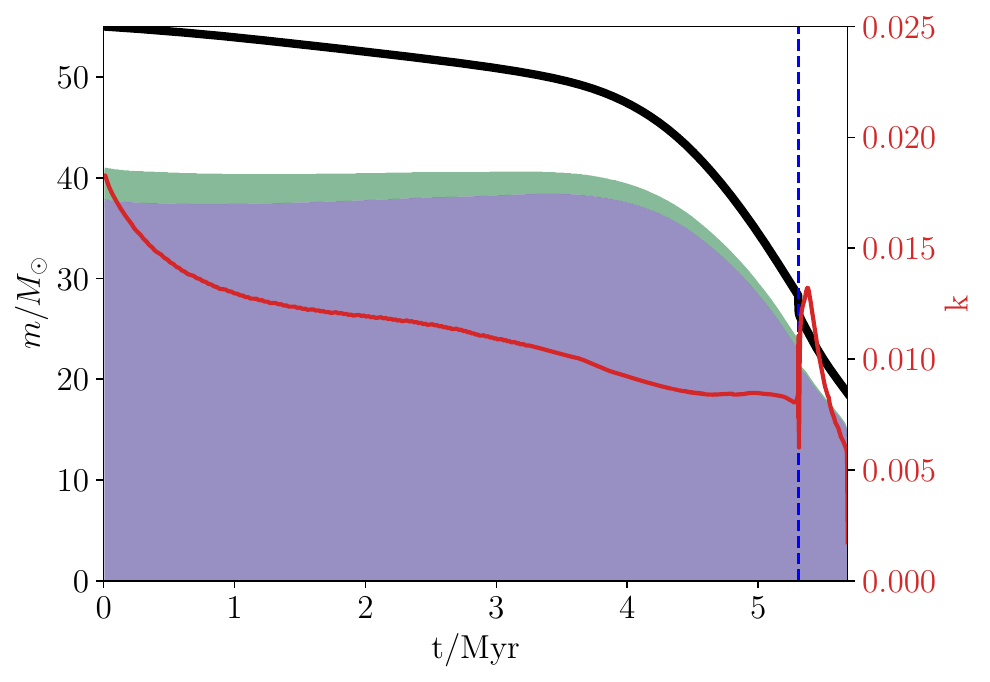}
\includegraphics[width=0.5\textwidth]{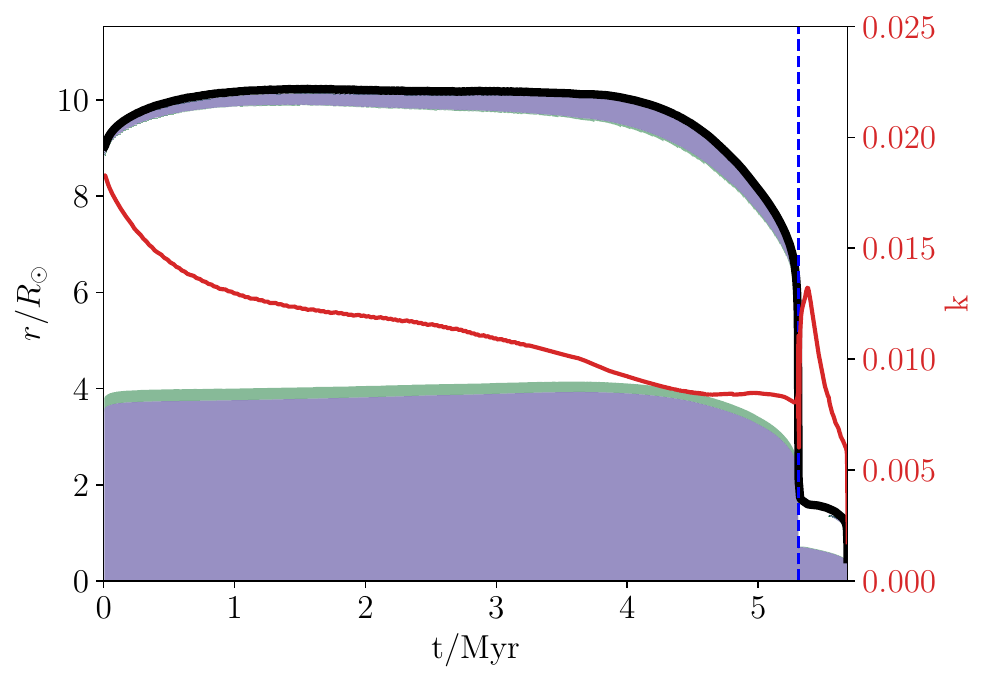}
    \caption{
    Same as Fig. \ref{fig:stellar_structure_low_Z} but for $Z=0.0042$.
    The vertical dashed blue  line  represents 5.31 Myr.
    }
    \label{fig:stellar_structure_high_Z}
\end{figure}

In our main analysis, we focussed on a default CHE binary at $Z=0.00042$ (Sect. \ref{sec:meth:binary_evolution}), which is 10\% of the default value used in \cite{2024ApJ...966....9S}. 
Here, we present the results of our CHE binary model at $Z=0.0042$.
Figure \ref{fig:stellar_structure_high_Z} illustrates the stellar structure. 
At the ZAMS, the total mass and the mass of the convective core are similar to those at the default metallicity. 
However, both the radius of the star ($\approx 10 R_{\odot}$) and the convective core ($\approx 4 R_{\odot}$) are slightly larger,which is also reflected in a lower value for the apsidal motion constant. 
Throughout its evolution, this stellar model loses significantly more mass compared to the default model, finishing its evolution with $\approx 18.6 M_{\odot}$. 
This is a direct consequence of line-driven winds, which are more efficient at higher metallicities. 
Mass loss through winds leads to a widening of the orbital period to $\approx 4.5$ d. 
If such a CHE binary were isolated and became a BBH, with negligible mass loss and change in the orbital period, it would take 25.5 Gyr to merge via gravitational-wave emission.

\begin{figure*}[]
    \centering
    \includegraphics[trim={0.7cm 8.5cm 0.7cm 9cm},clip,width=\columnwidth]{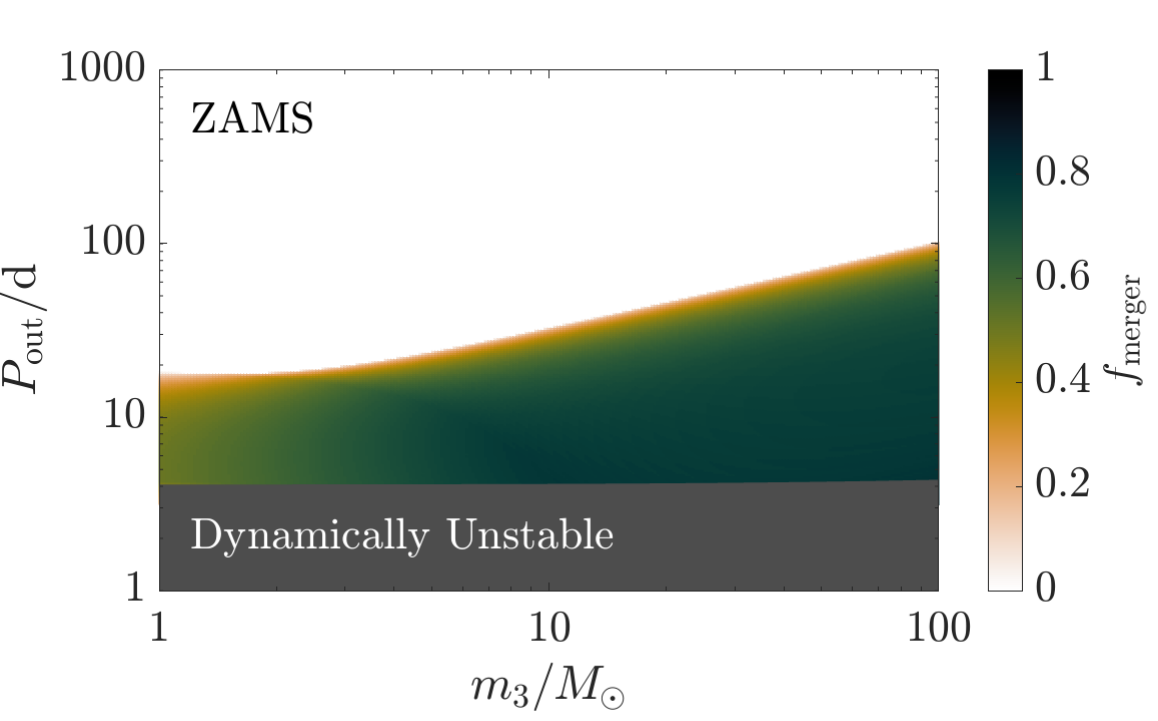}
    \includegraphics[trim={0.5cm 8.5cm 0.7cm 9cm},clip,width=\columnwidth]{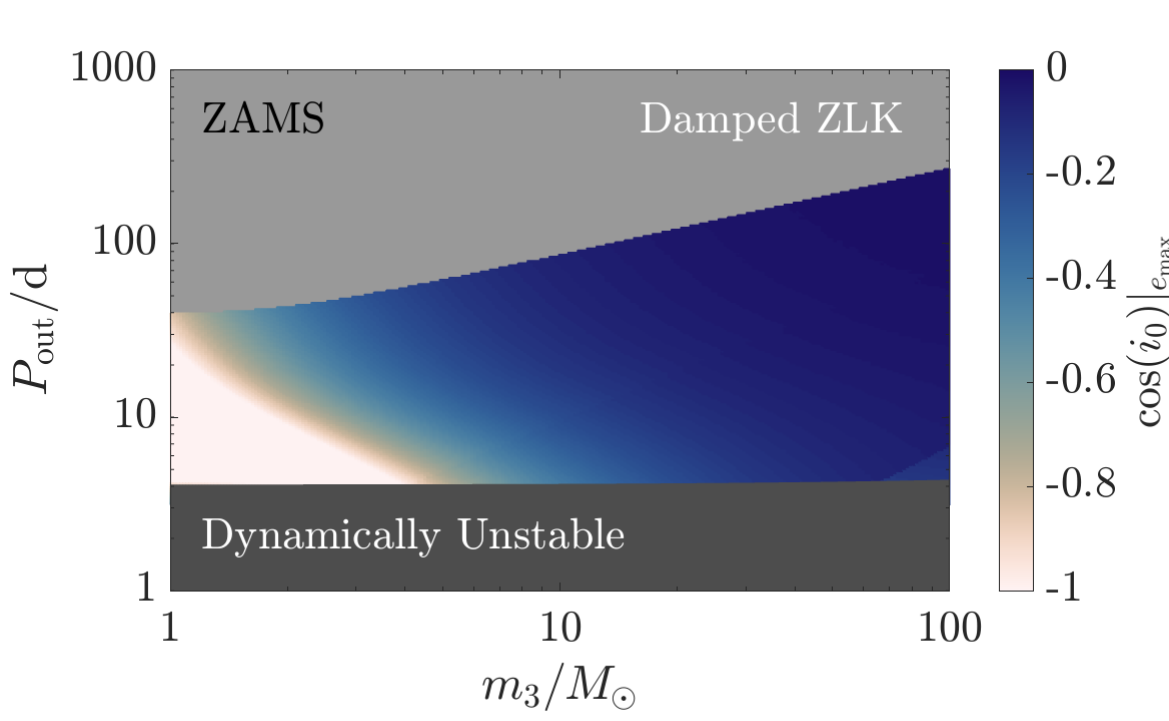}
    \includegraphics[trim={0.7cm 8.5cm 0.7cm 9cm},clip,width=\columnwidth]{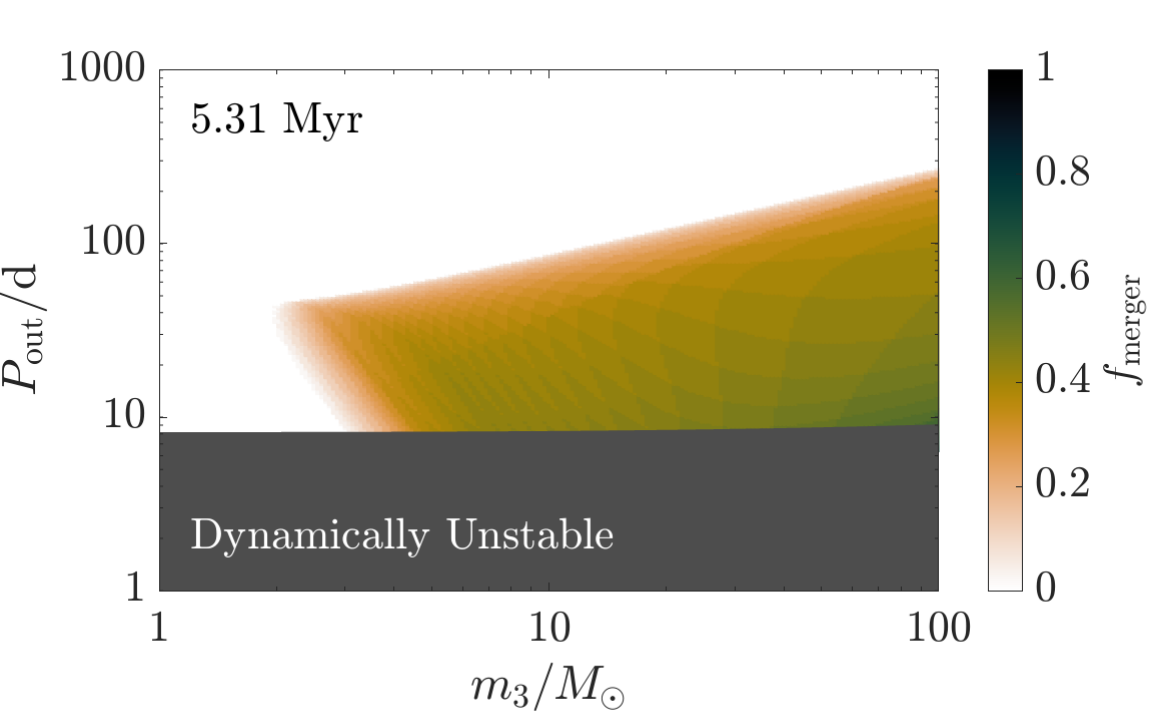}  
    \includegraphics[trim={0.5cm 8.5cm 0.7cm 9cm},clip,width=\columnwidth]{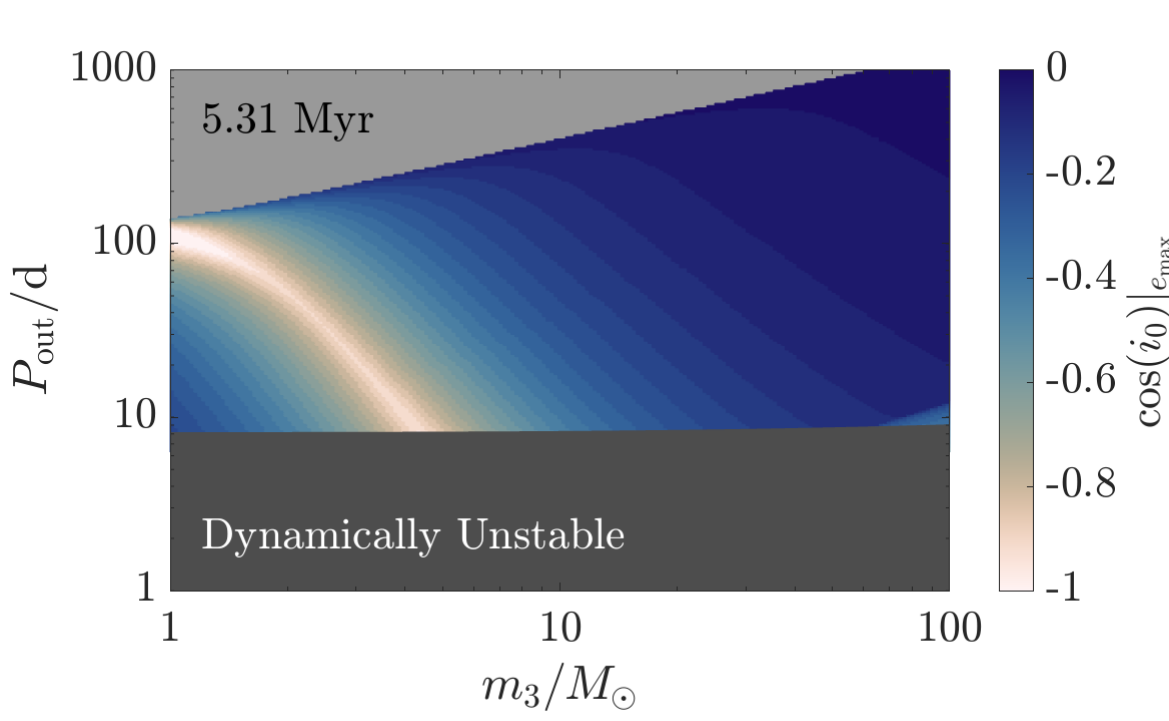}
    \caption{
    Same as Fig. \ref{fig:f_merger_CHE} but for $Z=0.0042$.
    }
    \label{fig:f_merger_CHE_high_Z}
\end{figure*}

Figure \ref{fig:f_merger_CHE_high_Z} illustrates the parameter space of an arbitrary tertiary system leading to stellar mergers. 
The results closely resemble those at lower metallicity, with a minor difference being a slightly broader parameter space in terms of the outer orbital period and the range of inclinations that meet our merger criteria. 
The primary effect of metallicity is shown in Fig. \ref{fig:summary_high_Z}, which demonstrates how wide inner binaries, which would not merge within the age of the Universe, can be dynamically assisted to merge once a black hole forms and tidal forces no longer apply.

\end{appendix}

\end{document}